\documentstyle[aps]{revtex}
\begin{document}
\newcommand{\beq}{\begin{equation}}
\newcommand{\eeq}{\end{equation}}
\newcommand{\beqn}{\begin{eqnarray}}
\newcommand{\eeqn}{\end{eqnarray}}
\newcommand{\bmath}{\begin{mathletters}}
\newcommand{\emath}{\end{mathletters}}
\twocolumn[\hsize\textwidth\columnwidth\hsize\csname @twocolumnfalse\endcsname 
\title{ Why holes are not like electrons:  A microscopic
           analysis of the differences between holes and electrons
           in condensed matter  }
\author{J. E. Hirsch }
\address{Department of Physics, University of California, San Diego\\
La Jolla, CA 92093-0319}

\date{\today} 
\maketitle 
\begin{abstract} 
We give a detailed microscopic analysis of why holes are different from electrons in
condensed matter. Starting from a single atom with zero, one and two electrons,
we show that the spectral functions for electrons and for holes are qualitatively
different because of electron-electron interactions. The quantitative importance
of this difference increases as the charge of the nucleus decreases. Extrapolating
our atomic analysis to the solid, we discuss the expected differences in the
single particle spectral function and in frequency dependent transport properties
for solids with nearly empty and nearly full electronic energy bands. We discuss the expected
dependence of these quantities on doping, and the physics of superconductivity
that results. We also discuss how these features of the atomic physics can be
modeled by a variety of model Hamiltonians.
\end{abstract}

\vskip2pc]
 
\section{Introduction}

The understanding of electronic correlation in solids and its consequences
for charge transport and for collective phenomena are fundamental problems 
in condensed
matter physics. At the simplest level, electronic correlation manifests
itself in the two-electron atom\cite{sla}. In this paper we return
to the old problem of the two-electron atom\cite{hyl} and show that a 
fundamental physical principle emerges from it, $electron-hole$
$asymmetry$. We then show that the understanding of electron-hole
asymmetry at the atomic level provides insight that implies that this asymmetry
has fundamental consequences for the physics of charge transport
and  collective phenomena of the solid state.

Remarkably, none of the many-body model Hamiltonians that are most widely used to study
the effect of electronic correlation in solids\cite{hubbard,hubbard2} such as the Hubbard model,
extended Hubbard model, degenerate Hubbard model, the Anderson impurity and lattice
models, the Kondo model, Falicov-Kimball model, 
Holstein model,  Su-Schrieffer-Heeger model and $t-J$ model, contain 
this very basic and fundamental aspect of electronic correlation 
that follows from the atomic analysis. Hence we argue that these
models are fundamentally flawed. We propose a variety of other model Hamiltonians
that contain this physics, which may be generically called 
'dynamic Hubbard models'\cite{spin,electron,polaron,dynhub}.
These models lead to a new understanding of the physics of charge transport
in solids, and to a universal mechanism of superconductivity\cite{hole1,hole2,undr}.

The observation that electrons and holes in atomic shells are in some sense equivalent 
was first made by Heisenberg\cite{heis}. It is easy to see that in the
absence of electron-electron interactions, an atom with $i$ electrons in an outer
partially filled shell defined by quantum numbers $n$, $l$, has the same multiplet
structure as an atom with $p-i$ electrons in this shell, with $p=4l+2$ the total number
of electrons that can fill this shell. In particular, this is trivially
true for an s-shell (l=0), where the state with one electron in the $ns^0$ shell
is $identical$ to the state with one hole in the $ns^2$ shell.
Furthermore, all the matrix elements of the Coulomb interaction operator,
both diagonal and off-diagonal, between these non-interacting electron atomic states,
are the same for a shell with $i$ electrons
and one with $i$ holes (i.e. $p-i$ electrons)\cite{bethe}. This of course
extends also to cases where more than one shell is incomplete. As a consequence,
the atomic multiplet structure of a shell with $i$ electrons is identical to the one
with  $i$ holes  within first order perturbation theory in the Coulomb
interaction between electrons. Thus, to the extent that first order 
perturbation theory holds for atoms, electron-hole symmetry exists for the
atomic  states, and
similar arguments can be used to argue for electron-hole symmetry in 
energy bands in the solid state\cite{ashcr}. This implies that the states of
atomic shells that are almost full, and of electronic energy bands that are almost
full, can be described in terms of the states of few positively charged
holes, instead of the states of many negatively charged electrons. The physical reality of the hole
concept for transport in solids was first demonstrated by Hall's measurement\cite{hall}
and Peierls' explanation\cite{peierls} of positive Hall coefficient 
in solids with nearly filled electronic energy bands.

However, in this paper we show that electron-hole symmetry in fact does not exist in
atoms, and consequently it will not exist in electronic energy bands in 
solids. Instead, we argue that electrons and holes are $fundamentally$ $different$
$objects$. In some sense, in the light of the above discussion, this can be understood 
as a failure of first order perturbation
theory. It is of course not surprising that perturbation theory in the
electron-electron Coulomb interaction should fail. The electronic stucture
of non-interacting electrons in atoms is determined by the Coulomb interaction 
between electrons and ions, which of course has the same coupling constant as the
electron-electron Coulomb interaction. Thus, it is $never$ true that 
the spacing between non-interacting electron atomic energy levels is larger than
matrix elements of the electron-electron Coulomb interaction, the regime
where perturbation theory may be expected to be valid. Despite of this,
it is fortunate that the qualitative structure of atomic multiplets and their
quantum numbers are the same for electron and holes\cite{heis}.

The physics of electron-hole asymmetry that results from electron-electron
interactions occurs even for $s$-shells, and even for the atomic $1s$ shell,
that is most isolated in energy. That is the case that we will analyze
in this paper. It is natural to expect that the physics that we find here
should have an even larger effect for atomic states with higher
quantum numbers, where shell occupation becomes larger and energy level
separation becomes smaller.

Since for an s-shell the one-electron state and the one-hole state are
the very same state, one may ask how it is possible to understand
electron-hole asymmetry based on it. The point is, in considering the
process of transport of charge in solids with electrons or holes, we need
to consider not only the  atom in a given charge state, but also
processes where electrons and holes are $created$ and $destroyed$ as
conduction occurs. It is these creation and destruction processes that
contain the physics of electron-hole asymmetry even at the level of
a single atom and even for s-shells.

This paper is organized as follows. In the next section we discuss the electronic
state of two electrons in the $1s$ shell of ions of nuclear charge $Z$. We show 
that the state of the two-electron atom is strongly influenced by electronic
correlation effects, the more so the smaller $Z$ is. In section III we discuss
the calculation of the atomic spectral function for electrons and holes, 
and the qualitative reasons for why they are fundamentally different.
Section IV discusses results for the atomic spectral function within the various
approximations for the state of the two-electron atom reviewed in Sect. II. In Sect. V 
we discuss the consequences of these results for the properties of electrons and
holes in the solid state, in particular for the single electron spectral
function and for frequency-dependent transport. In section VI we discuss
 the consequences
of this physics for the understanding of superconductivity in solids. Sect. VII  
discusses several model Hamiltonians that contain the physics of electron-hole
asymmetry found at the atomic level.
We conclude in Sect. VIII with a summary and discussion of our results, a review 
of the empirical and experimental evidence in support of this physics as the 
underlying universal mechanism of superconductivity in solids, and a survey of
some of the many open questions and opportunities for further research in this area.

\section{The two-electron atom}

The wavefunction for an electron in the lowest energy state ($1s$) of a 
hydrogen-like ion of nuclear charge $Z$ is
\beq
\varphi _{Z}(r)=(\frac{Z^3}{\pi})^{1/2} e^{-Zr}
\eeq
with $r$ measured in units of the Bohr radius $a_0$. The ground state wave function
of the two-electron ion is $not$, of course,
\beq
\Psi(r_1,r_2)=\varphi_Z(r_1)\varphi_Z(r_2)
\eeq
because of electron-electron interactions. Consider the following approximations to
the ground state of the two-electron ion:
\subsection{The Hartree wavefunction}
The simplest approximate wavefunction that takes into account the effect of
electron-electron interaction is of the Hartree form
\bmath
\beq
\Psi_H(r_1,r_2)=\varphi_{\bar{Z}} (r_1)\varphi_{\bar{Z}} (r_2)
\eeq
where
\beq
\bar{Z}=Z-\frac{5}{16}
\eeq
\emath
results from minimization of the total energy for a variational wave function
of the form Eq. (3a). Eq. (3) describes an expanded orbital for each electron,
due to the partial shielding of the positive nuclear charge by the other
electron. In Eq. (3), both electrons reside in the same expanded orbital.
A slightly better approximation of the Hartree form is obtained by allowing for the
most general single particle wave function rather than the exponential form
Eq. (1). For example, the error in the ground state energy of the $He$ atom is
$1.95\%$ with Eq. (3) and $1.45\%$ with the optimal single particle wavefunction
in the Hartree wavefunction\cite{sla}.

\subsection{The Eckart wavefunction}
A better approximation is obtained by allowing for radial correlations, through
the wave function\cite{eck}
\bmath
\beq
\Psi_E(r_1,r_2)=\frac{\varphi_{Z_1} (r_1)\varphi_{Z_2} (r_2)
+\varphi_{Z_2} (r_1)\varphi_{Z_1} (r_2)}
{2(1+S_{12}^2))^{1/2}}
\eeq
\beq
S_{12}=(\varphi_{Z_1},\varphi_{Z_2})
\eeq
\emath
where the exponents $Z_1$ and $Z_2$ are again obtained by minimization of the
total energy, for a wavefunction of the form Eq. (4). The numerical results for
$Z_1$ and $Z_2$ obey the approximate relations
\bmath
\beq
Z_1=1.14Z-0.105
\eeq
\beq
Z_2=0.905Z-0.622.
\eeq
\emath
Hence, one of the electrons resides in an orbit of approximately the same radius
as that of the single electron ion, and the second one resides in a substantially
enlarged orbit. The minimization procedure resulting in the values of
$Z_1$ and $Z_2$ becomes unstable for $Z\leq 0.93$. For $He$, the error in the
ground state energy is now reduced to $0.98\%$.

\subsection{The Hylleraas wavefunction}
Much more accurate wavefunctions for the two-electron system are obtained
by introducing dependence of the wavefunction on $r_{12}=|\vec{r}_1-\vec{r}_2|$
in addition to $r_1$ and $r_2$, which allows for angular in addition to
radial correlations. We consider here the simplest wavefunction of this
form, Hylleraas' ``dritte Naherung'' (third approximation)\cite{hyl}:
\bmath
\beq
\Psi_{Hy}(\vec{r}_1,\vec{r}_2)=N\varphi(2Zks,2Zkt,2Zku)
\eeq
\beq
\varphi(s,t,u)=e^{-s/2}[1+c_1u+c_2t^2]\eeq
\beq
s=r_1+r_2
\eeq
\beq
t=r_2-r_1
\eeq
\beq
u=r_{12}
\eeq
\emath
The parameters $c_1$, $c_2$ and $k$ are determined by minimization of the
energy, as described in the appendix, and $N$ is a normalization constant. 
Both this as well as the previous 
wave functions yield upper bounds to the ground state energy. $c_1$ and $c_2$ are
found to be positive, indicating that for given $s=r_1+r_2$ the wavefunction
is larger for large  angle between $\vec{r}_1$ and $\vec{r}_2$ (angular correlations)
and for large $|r_1-r_2|$ (radial correlations). The error in the $He$ energy with the
wavefunction Eq. (6) is reduced to $0.05\%$. Table 1 summarizes the values of
the energy and orbital exponents for the three wavefunctions discussed for the
$He$ ($Z=2$) and the $H^-$ ($Z=1$) ions compared to the experimental values.

Figure 1 shows the orbital exponents versus ionic charge in the different
approximations. The Hylleraas wave function has an orbital exponent similar
to the Hartree wave function, while the exponents of the Eckart wave functions are very 
different. This suggests that even though the Eckart wave function gives better 
results than the Hartree one for the energy, the Hartree method may give
a better representation of the wavefunction itself. Figure 2 shows the parameters 
$c_1$ and $c_2$ of the Hylleraas wavefunction, which describe angular and radial correlations
respectively, versus ionic charge. Note that both parameters increase rapidly as
the ionic charge decreases, and that angular correlations are much more important
than radial correlations for large $Z$.

The different wavefunctions discussed above describe attempts of the electronic 
wavefunction to reduce the Coulomb repulsion between electrons, without paying unduly
in electron-ion energy. The effective Coulomb repulsion between electrons is
\beq
U_{eff}(Z)=E(2)+E(0)-2E(1)
\eeq
with $E(n)$ the ground state energy for the ion with $n$ electrons 
($E(0)=0$, $E(1)=-Z^2$, with $E$ in $Ry=13.6eV$). For the wavefunction 
Eq. (2) with the electrons occupying the same orbitals as in the singly
occupied atom
\bmath
\beq
U_{eff}(Z)\equiv U=\frac{5}{4}Z
\eeq
\beq
U=\int d^3r d^3 r' |\varphi _{Z}(r)|^2 \frac{e^2}{|r-r'|}|\varphi _{Z}(r')|^2
\eeq
\emath
and for the Hartree wave function Eq. (3)
\beq
U_{eff}^H=\frac{5}{4}Z-\frac{25}{128},
\eeq
smaller than Eq. (8a) because the wavefunctions are more expanded
(note however that this is different from the actual Coulomb integral
Eq. (8b) computed with the Hartree wavefunctions, due to the cost in
electron-ion energy payed by the expanded wave functions). For the
Eckart and Hylleraas wavefunctions the effective $U$ becomes
progressively smaller, corresponding to the decrease of the
two-electron energy $E(2)$ with increasingly better wavefunctions.
Figure 3 shows $U_{eff}$ versus ionic charge in the different approximations.

Figure 4 depicts qualitatively the electrons in the atom in the different
approximations. We emphasize that the two-electron wavefunction in any of the
approximations discussed is very different from the one corresponding to
non-interacting electrons, Eq. (2). For the purposes of this paper, any
of these wavefunctions, including the simple Hartree one, describes
'electronic correlation', in the sense that the two-electron wavefunction
is different from the one given in Eq. (2), and is appropriate to
illustrate the physics of electron-hole asymmetry. Note also that the
difference between these wave functions and the uncorrelated wavefunction
Eq. (2) increases as the ionic charge $Z$ decreases and the relative
importance of electron-electron versus electron-ion interaction strength
increases.

It is also worth pointing out that the electron-electron repulsion $U_{eff}$ 
decreases as the ionic charge $Z$ decreases for all these wavefunctions,
including the uncorrelated one. Hence, everything else being equal, pairing
of electrons (or holes) should be favored when the effective ionic charge is small,
i.e. with negatively charged ions. This simple fact, which provides some
rationale for high temperature superconductivity being favored in systems with
negative ions ($O^=$ in cuprates, $B^-$ in $MgB_2$) has surprisingly not
been pointed out before to our knowledge.

\section{Atomic spectral functions}

Properties of many-electron systems can be studied by consideration of 
spectral functions. The zero temperature one-electron spectral function
for a many-body system is conventionally defined as 
\bmath
\beqn
A(\omega)&=&\sum_{n=1}^\infty [|<n|c_{\alpha\sigma}^\dagger|1>|^2\delta
(\omega-(E_n^{N+1}-E_1^{N+1}+\mu_N)) \\ \nonumber &+&
|<n|c_{\alpha\sigma}|1>|^2\delta(\omega+(E_n^{N-1}-E_1^{N-1}-\mu_{N-1}))].
\eeqn
where $|1>$ is the ground state of the $N$-electron system,  $E_1^N$ its
ground state energy, and $\mu_N=E_1^{N+1}-E_1^N$, $\mu_{N-1}=E_1^{N}-E_1^{N-1}$.
For a metal, $\mu_N=\mu_{N-1}=\mu$, and one can redefine the frequency
$\omega \rightarrow \omega + \mu$ so that Eq. (10a) becomes
\beqn
A(\omega)&=&\sum_{n=1}^\infty [|<n|c_{\alpha\sigma}^\dagger|1>|^2\delta
(\omega-(E_n^{N+1}-E_1^{N+1})) \\ \nonumber &+&
|<n|c_{\alpha\sigma}|1>|^2\delta(\omega+(E_n^{N-1}-E_1^{N-1}))].
\eeqn
\emath
and the zeros of the $\delta$-function arguments 
occur at the excitation energies of the system.
To have a smooth transition between the metal and atomic spectral functions
we will here define the atomic spectral functions as Eq. (10b) rather than
Eq. (10a).
The first term describes the response of the system when an electron of spin 
$\sigma$ is created in the
single-particle state $\alpha$, into the ground state with $N$ electrons.
$E_n^{N+1}$ is the n-th excited state energy of the many-electron system
with $N+1$ electrons. Similarly, the second term describes the response
of the system when an electron of spin $\sigma$ at the single-particle state $\alpha$ is
destroyed from the ground state of the $N$-electron system, and 
$E_n^{N-1}$ are the excited state energies of the resulting $(N-1)$-electron
system. Generically, in many-body systems the one-electron spectral function
is of the form\cite{mahan}
\beq
A(\omega)=z\delta(\omega-(\epsilon-\mu))+A_{inc}(\omega)
\eeq
where the first term describes the quasiparticle, with quasiparticle
weight $z$, $0\leq z \leq 1$, and the second term describes a 
continuum of incoherent excitations at higher energies. For a small system
like an atom however, the spectral function will consist of only discrete 
$\delta$-functions. Nevertheless, we can identify the lowest-energy 
$\delta$-function as corresponding to the quasiparticle, and its coefficient
as the quasiparticle weight.

The fundamental asymmetry between electrons and holes follows immediately
from consideration of these spectral functions. Consider the spectral
function for creating an electron in the empty atom, $N=0$. From Eq. (10b),
it is simply
\beq
A_{el}(\omega)=\delta(\omega)
\eeq
so that it is totally coherent, with quasiparticle weight $z=1$.

Consider next the spectral function for creating a hole in the full
$1s$ shell:
\bmath
\beq
A_h(\omega)=\sum_n|<n|c_{\alpha,\uparrow}|1>|^2\delta(\omega+E_n^1-E_1^1)
\eeq
The excitation energies in Eq. (13a) are the ones of the singly-occupied
atom,
\beq
E_n^1=-\frac{Z^2}{n^2}
\eeq
and the matrix elements are 
\beq
<n|c_{\alpha,\uparrow}|1> \equiv S_n  =\int d^3r_1 d^3r_2 \Psi(\vec{r}_1,\vec{r}_2)
\varphi_\alpha(r_1)\varphi_n(r_2)
\eeq
\emath
with $\Psi$ the two-electron ground state wave function, $\varphi_n$ the
wavefunction for the n-th excited state of the singly occupied atom, and
$\varphi_\alpha$ the single particle wave function of state $\alpha$. The important
qualitative point is, because $\Psi$ is not given by the product form
of single electron wavefunctions Eq. (2), the spectral function
Eq. (13a) $necessarily$ consists of a $sum$ of $\delta$-functions rather 
than a single one as Eq. (12):
\bmath
\beq
A_h(\omega)=z\delta(\omega)+\sum_{n\neq 1} S_n^2 \delta(\omega+E_n^1-E_1^1)
\eeq
\beq
z=S_1^2
\eeq
\emath
Hence, the spectral function for hole creation Eq. (14a) is
$qualitatively$ $different$ from the one for electron
creation Eq. (12). 

The foregoing remarks are true for $any$ single-particle wave function
$\varphi_\alpha$. Which is the appropriate $\varphi_\alpha$ to use
to define the hole spectral function? We argue that it is reasonable to use
the single particle wave function that maximizes the quasiparticle weight
of the hole spectral function, $S_1^2$ (such a hole wavefunction will have
minimal kinetic energy in the solid). For the Hartree wavefunction
Eq. (3a) it is simply given by
\bmath
\beq
\varphi_\alpha(r)=\varphi_{\bar{Z}}(r)
\eeq
and
\beq
S_1=\int d^3r \varphi_1(r)\varphi_{\bar{Z}}(r)=\frac{(Z\bar{Z})^{3/2}}
{(\frac{Z+\bar{Z}}{2})^3}
\eeq
\emath
with $\bar{Z}=Z-5/16$. For the Eckart and Hylleraas wave functions, we use
a single particle wave function of the form Eq. (15a), with $\bar{Z}$ determined
by maximization of the function
\beq
S_1(\bar{Z})=\int d^3r_1d^3r_2 \Psi(\vec{r}_1,\vec{r}_2)
\varphi_{\bar{Z}}(r_1)\varphi_1(r_2)
\eeq

Note that the hole spectral function as defined by Eq. (13) satisfies
\beq
\int_0^\infty d\omega A_h(\omega)=\sum_n S_n^2=
\int d^3r_1 [\int d^3r_2 \Psi(r_1,r_2)\varphi_\alpha(r_2)]^2
\eeq
In the Hartree approximation
\beq
\int_0^\infty d\omega A_h(\omega)=1   ,
\eeq
while with the more accurate wave functions, the frequency integral of the
hole spectral function is less than unity. This is because there is also
a non-zero probability of creating an $electron$ in the single particle state
$\varphi_\alpha$ into the two-electron atom. However that probability
is found to be small.

Finally, the atomic spectral function for the singly occupied atom is of
interest. We consider separately the positive and negative frequency parts. The
spectral function for destruction of an electron in the singly-occupied
atom is:
\beq
A'_{el}(\omega) = \sum_n |<n|c_{\alpha \uparrow}|1>|^2\delta(\omega+E_n^0-E_1^0)
\eeq
which is of course also given by
\beq
A'_{el}(\omega)=\delta(\omega)
\eeq
and the spectral function for destruction of a hole in the singly occupied atom,
or equivalently creation of an electron in the singly occupied atom, is :
\beq
A'_h(\omega)=\sum_n (S'_n)^2\delta(\omega-(E_n^2-E_1^2))
\eeq
where $E_n^2$ are the excitation energies of the doubly occupied atom, and
\beq
S'_n=\int d^3 r_1 d^3 r_2 \varphi_\alpha(r_1)\varphi_Z(r_2)\Psi_n(r_1,r_2)
\eeq
with $\Psi_n$ the wavefunctions for the excited states of the
two-electron atom. Once again, the spectral function $A_h'$ is qualitatively
different from $A'_{el}$, as it contains an incoherent part.
The spectral function for the singly occupied atom as defined by Eq. (10b) 
will be either Eq. (20) or Eq. (21) depending on the value of $\sigma$ in
Eq. (10b) and of the spin of the electron
in the singly occupied atom .

To estimate $S_n'$ we need the excited state wavefunctions  for the
doubly occupied ion. The excited states are approximately described by one electron
in the ground state and the other electron in an excited state of the ion with
orbital exponent $\tilde{Z}$, $\varphi_{n\tilde{Z}}$, with $\tilde{Z}$
 given by Slater's rules\cite{sla}:
\bmath
\beq
\tilde{Z}=Z-0.85 , n=2
\eeq
\beq
\tilde{Z}=Z-1 , n=3,4,...
\eeq
\emath
The matrix elements are then given by (for $n>1$):
\beq
S_n'=\int d^3r \varphi_\alpha (r)\varphi_{n\tilde{Z}}(r)
\eeq
and the excitation energies by
\beq
E^2_n=-Z^2-\frac{\tilde{Z}^2}{n^2} .
\eeq

\section{Results for the atomic spectral functions}

As discussed in the previous section, we calculate the hole spectral function for that
single particle wave function $\varphi_\alpha$ that maximizes the
quasiparticle weight, which we will call the 'hole wavefunction'. Figure 5 shows
the orbital exponent of that state, $\alpha$, versus $Z$. For the Hartree
approximation $\alpha=\bar{Z}$, for the Eckart and Hylleraas approximations
$\alpha$ is somewhat smaller, i.e. the hole wavefunction is somewhat
more extended. In all approximations the hole wavefunction becomes more
diffuse as the ionic  charge $Z$ decreases, as expected. Figure 6 shows the
quasiparticle weight for the hole, from Eqs. (14b) and (16). The Hartree
wavefunction somewhat overestimates and the Eckart wavefunction somewhat
underestimates the quasiparticle weight given by the Hylleraas wavefunction, which
presumably is the most accurate one. As the ionic charge decreases the 
quasiparticle weight for the hole decreases in all the approximations.

To obtain the full spectral function for hole creation we compute the matrix
elements $S_n$, Eq. (13c). Because $\varphi_\alpha$ is spherically symmetric,
only the excited atomic s-states give non-zero results for the integral. They are
given by
\bmath
\beq
\varphi_n (r)=(\frac{(Z/n)^3}{\pi})^{1/2}
\sum_{l=0}^{n-1} (-1)^l a_l (\frac{Zr}{n})^l e^{-Zr/n}
\eeq
\beq
a_l=\frac{(n-1)! 2^l}{(n-l-1)!(l+1)! l!}
\eeq
\emath
The integrals for all the approximate wavefunctions considered can be done
analytically, as described in Appendix B.

Figure 7 shows the spectral function for hole creation in the full $1s$ shell,
Eq. (13a), for various values of $Z$. We show results for both the Hartree
and Hylleraas wavefunctions; the results for the Eckart wave function are
similar. For $Z=2$, $A_h(\omega)$ is close to a $\delta-$function at $\omega=0$, of weight
unity; the incoherent part occurs at very high energies and has very small weight.
As $Z$ decreases, the weight of the $\omega=0$ peak (quasiparticle) gradually
decreases, and the weight of the higher enery peaks (incoherent part) gradually
increases; in addition, the incoherent peaks shift to lower frequency.
In Fig. 8 we show results with the Hartree wavefunction for an even smaller
$Z$, $Z=0.4$ (the procedure to obtain the Hylleraas wavefunction does not converge
for such a small $Z$), showing that here the incoherent contribution is
bigger than the quasiparticle part.

Similarly we can calculate the spectral function for hole destruction
(electron creation) in the ion with one hole (i.e. with one electron),
from Eqs. (23)-(25). The excited states wavefunctions are given by Eq. (26)
with $Z$ replaced by $\tilde{Z}$ given in Eq. (23). The matrix elements
turn out to be similar to the case of hole creation, but the excitation energies
here are much lower. Hence the incoherent part of the spectral function is
shifted to much lower energies for these processes. Figure 9 shows the calculated
spectral function for hole destruction for $Z=2$ and $Z=1$.

Note that it is meaningless to calculate the spectral function for hole destruction
for the isolated atom for $Z<1$. In that case, all the excited states of the
two-electron ion are unbound. However the situation is different in the solid state.
There, the negative ion (e.g. $O^=$ or $B^-$) will be surrounded by
positive cations, and excited states can be formed where the excited electron
is still confined in the neighborhood of the anion and the surrounding cations.

\section{Consequences for the solid state}

We have seen in the last section that the spectral functions for electrons and
holes in the atom will be either single $\delta$-functions or sums of 
$\delta$-functions depending on the charge state of the atom. In this section
we discuss the fundamental consequences of this for the solid state.

When we bring together the atoms to form a solid, orbitals overlap and
bands are formed. Consider for simplicity the lowest band, formed by
overlaping $1s$ orbitals, well separated from other bands if the interatomic
distance is large. The kinetic energy operator for an electron in such a
band, in second quantized formalism, is
\bmath
\beq
H=-\sum_{i,j} t_{ij} c^\dagger_{i\sigma}c_{j\sigma}
\eeq
where $t_{ij}$ is the Fourier transform of the Bloch band energy
\beq
t_{ij}=\frac{1}{N}\sum_{i,j}e^{ik(R_j-R_i)}\epsilon_k
\eeq
\emath
and $c^\dagger_{i\sigma}$ creates an electron in Wannier orbital $i$,
a linear combination of nearby atomic orbitals with largest amplitude
at site $i$. In the simplest tight binding approximation the
hopping matrix element for a single electron can be obtained from
\bmath
\beq
t_{ij}=\frac{-(\varphi_i,h\varphi_j)+S_{ij}(\varphi,h\varphi_i)}{1-S_{ij}^2}
\eeq
\beq
S_{ij}=(\varphi_i,\varphi_j)
\eeq
\emath
with
\beq
h=-\nabla ^2 -\frac{2Z}{|r-R_i|}-\frac{2Z}{|r-R_j|}
\eeq
the single electron Hamiltonian (in atomic units). Eq. (28) reproduces quite
accurately the spacing between bonding and antibonding states of the
single electron diatomic molecular ion in a wide range of interatomic
distances and ionic charges\cite{molec}. However we emphasize that Eq. (27) is generally
valid for any Bloch band: away from the tight binding limit the hopping
$t_{ij}$ will involve many neighboring atoms and differ from Eq. (28), and the
Wannier orbital will be a linear combination of many atomic orbitals. We will focus
on the tight binding limit here from simplicity, but expect the qualitative 
physics to survive beyond that limit.

\subsection{Spectral function}
\subsubsection{General considerations}

Consider the spectral function for electrons in this band, $A(k,\omega)$. For
the empty band, we use Eq. (10) with $c_{\alpha\sigma}$ replaced by
\beq
c_{k\sigma}=\frac{1}{\sqrt{N}}\sum_ie^{ikR_i}c_{i\sigma}.
\eeq
When $c_{k\sigma}^\dagger$ acts on the ground state (the empty band), it 
yields a single electron Bloch function, an eigenstate of the Hamiltonian
Eq. (27a). Hence the spectral function is simply given by
\bmath
\beq
A(k,\omega)=\delta(\omega-(\epsilon_k-\mu))
\eeq
\beq
\epsilon_k=\frac{1}{N}\sum_{i,j}e^{ik(R_i-R_j)}t_{ij}
\eeq
\emath
with $\mu$ the chemical potential (=the bottom of the band for the empty band).

For  the full band instead, the situation is $much$ more
complicated. We wish to compute Eq. (10) with the operator
\beq
c^\alpha_{k\sigma}=\frac{1}{\sqrt{N}}\sum_ie^{ikR_i}c^\alpha_{i\sigma}
\eeq
where the atomic destruction operator $c_{i\sigma}^\alpha$ destroys an electron
in the single electron atomic state $\alpha$ with wavefunction $\varphi_\alpha$
as discussed in the previous section. The ground state $|0>$ is given by
\beq
|0>=|\uparrow\downarrow>_1 |\uparrow\downarrow>_2 ... |\uparrow\downarrow>_i...
|\uparrow\downarrow>_N
\eeq
where $|\uparrow\downarrow>_i$ is the correlated two-electron ground state
of the i-th atom, described e.g. by the Hylleraas wavefunction. When
$c_{i\sigma}^\alpha$ operates on this state we obtain
\beq
c_{i\sigma}^\alpha |\uparrow\downarrow>_i=S_1 |\downarrow>_i+
\sum_{n>1} S_n |\downarrow^n>_i
\eeq
where the matrix elements $S_n$ are given by Eq. (13c) and (B7), and $|\downarrow^n>_i$
denotes the n-th excited state of the single electron ion with energy given by
Eq. (13b), and $|\downarrow>\equiv |\downarrow^1>$. Hence,
\beq
c_{k\sigma}^\alpha |0>=\frac{S_1}{\sqrt{N}} \sum_i e^{ikR_i} |\downarrow>_i+
\frac{1}{\sqrt{N}}\sum_i e^{ikR_i}\sum_{n>1} S_n|\downarrow^n>_i   
\eeq
which is, not surprisingly, $not$ an eigenstate of the Hamiltonian Eq. (27a).
(In Eq. (35), the state vectors for all sites $j\neq i$, $|\uparrow\downarrow>_j$, are
omitted for clarity.) Thus the spectral function for a single hole in the full band
is $not$ a single $\delta-$function and hence is qualitatively different from 
the spectral function of a single electron in 
the empty band, similarly to the case for the single atom. 

The first term in Eq. (35) denotes a Bloch wave for a
single hole in the full band. This is also not an exact eigenstate of the Hamiltonian
Eq. (27a), because every time the hole hops to a neighboring site there is
a finite probability amplitude that the final state has the atoms in excited
states, as discussed in the previous section. Nevertheless, within the
approximation where those processes are excluded, the energy of the itinerant hole
will be given by
\beq
\epsilon_k^h=S_1^2\epsilon_k=z\epsilon_k\equiv\tilde{\epsilon}_k .
\eeq
Separating this contribution from the rest we have for the single hole spectral
function
\beq
A_h(k,\omega)=z\delta(\omega-(\tilde{\epsilon}_k-\mu))+A'(k,\omega)
\eeq
where $A'(k,\omega)$ contains all contributions to the spectral function with
atomic excited states resulting from the second term in Eq. (35). Hence it will 
involve the high energy part of the atomic spectral function discussed in the
previous sections. In a solid, we expect that this contribution will not have a strong
k-dependence and that the atomic $\delta$-functions will broaden to give
rise to the 'incoherent' part of the spectral function, $A'(k,\omega)$.

The derivation of Eq. (37) is not rigurous, and hence the expressions for
$z$ and $\tilde{\epsilon}_k$ are not exact. Nevertheless, the general
form Eq. (37), possibly with some $k$-dependence to the quasiparticle weight
$z$, is expected to be correct for a many-body system\cite{mahan}. In fact, Sham
has shown rigurously\cite{sham} that the spectral function for 
a sigle hole in a full band has a $\delta$-function quasiparticle contribution
that is separated from the continuum of incoherent excitations. We believe that
the k-independent $z$ and the form of $\tilde{\epsilon}_k$ obtained
are reasonable approximations at least in the tight binding limit. In particular, this 
implies that the quasiparticle
weight $z$ and the effective mass renormalization for a hole
\beq
\frac{m^*}{m}=\frac{1}{z}
\eeq
will have the qualitative dependence on the ionic charge $Z$ discussed in the
previous section.

\subsubsection{Quasiparticle operators}

In the foregoing discussion we considered the spectral function for a single electron in
the empty band and for a single hole in the full band.
To generalize this analysis to other band fillings we define quasiparticle operators.
The atomic electron operator can be written in terms of atomic states as
\beq
c_{i\uparrow}=|0><\uparrow|+S_1 |\downarrow><\uparrow\downarrow|+
\sum_{(n,n')\neq (0,0)} S_{n n'} |\downarrow^n><\uparrow\downarrow^{n'}|
\eeq
where $|\downarrow^n>$ and $|\uparrow\downarrow^{n'}>$ are excited states of
the singly and doubly occupied ion respectively, and 
\beq
S_{n n'}=\int d^3r d^3 r' \Psi_{n'}(r_1,r_2)\varphi_\alpha(r_1)\varphi_n(r_2)
\eeq
with $S_{n 1}=S_n$, $S_{1 n'}=S'_{n'}$ as discussed in the previous section.
We write Eq. (39) as
\beq
c_{i\sigma}=[1+(S_1-1)\tilde{n}_{i,-\sigma}]\tilde{c}_{i\sigma}
+c'_{i\sigma}
\eeq
where the 'quasiparticle operator' $\tilde{c}_{i\sigma}$ acts as if 
electronic correlations didn't exist:
\bmath
\beq
\tilde{c}_{i\sigma}^\dagger |0>=|\sigma>
\eeq
\beq
\tilde{c}_{i\sigma} |-\sigma>=|\uparrow\downarrow>
\eeq
\emath
and obeys usual anticommutation relations. 
$\tilde{n}_{i\sigma}=\tilde{c}_{i\sigma}^\dagger \tilde{c}_{i\sigma}$. The operator
$c'_{i\sigma}$ is the 'incoherent part' of the bare fermion operator $c_{i\sigma}$,
given by the last term in Eq. (39). From Eq. (41), we obtain that the quasiparticle weight 
for a band with band filling of n electrons per atom is approximately given by
\beq
z(n)=[1+(S_1-1)\frac{n}{2}]^2   
\eeq
decreasing monotonically from $1$ to $S_1^2$ as $n$ increases from
$0$ to $2$.

\subsubsection{Quasiparticle Hamiltonian}

The effective low energy Hamiltonian for quasiparticles (kinetic energy part only)
is obtained by replacing
the bare fermion operators in Eq. (27a) in terms of their form Eq. (41) and
neglecting the 'incoherent part' of the operators that would leave the
atoms in excited states. Hence, it describes only ground-state to
ground-state transitions (diagonal transitions in the language of small polaron
theory)
\bmath
\beq
H_{qp}=-\sum_{ij}t_{ij}^\sigma \tilde{c}_{i\sigma}^\dagger\tilde{c}_{j\sigma}
\eeq
\beqn
t_{ij}^\sigma&=&t_{ij}[1-(1-S_1)(\tilde{n}_{i,-\sigma}+\tilde{n}_{j,-\sigma})\nonumber \\
& &+(1-S_1)^2\tilde{n}_{i,-\sigma}\tilde{n}_{j,-\sigma}] .
\eeqn
\emath
In a mean field approximation, the quasiparticle energy is simply
given by taking the expectation value of Eq. (44b), yielding
\beq
\tilde{\epsilon}_k=z(n)\epsilon_k .
\eeq
To simplify the notation, we omit to denote explicitely the $n$-dependence
of $\tilde{\epsilon}_k$ on the left side of Eq. (45). Eq. (45) implies
that the relation Eq. (38) between effective mass and quasiparticle weight
is valid for any band filling, and that the quasiparticle mass increases and
the effective bandwidth decreases with increasing band filling.

\subsubsection{Phenomenological model for the spectral function}

From the considerations in the previous sections we conclude that a reasonable
model for the spectral function in the solid that contains the physics
discussed is of the form
\beq
A(k,\omega)=z(n)\delta(\omega-(\tilde{\epsilon}_k-\mu))+A'(\omega)
\eeq
where $A'(\omega)$ is the incoherent part describing the atomic
excitations, and we have ignored its k-dependence. The quasiparticle weight
$z(n)$ and energy $\tilde{\epsilon}_k$ 
are given by Eqs. (43) and (45) respectively. The atomic spectral functions
discussed in Sect. III will be broadened in the solid state, and we
assume the Gaussian forms
\bmath
\beq
A_h(\omega)=\frac{1}{\sqrt{2\pi}\Gamma_1} 
e^{-\frac{(\omega+\epsilon_1)^2}{2\Gamma_1^2}}
\eeq
\beq
A_h'(\omega)=\frac{1}{\sqrt{2\pi}\Gamma_2} 
e^{-\frac{(\omega-\epsilon_2)^2}{2\Gamma_2^2}}
\eeq
\emath
corresponding to hole creation  (negative $\omega$) and hole destruction
(positive $\omega$) respectively. 
$\epsilon_1$ and $\epsilon_2$ are mean excitation energies of the
singly occupied ion and the doubly occupied ion, hence $\epsilon_1>\epsilon_2$,
and we assume the broadening $\Gamma_i$ is proportional to $\epsilon_i$. The
relative weights of $A_h$ and $A_h'$ in $A'$ will be determined by the
probabilities of  doubly and singly ocuppied sites respectively  in the solid for band
filling $n$. The probabilities for empty, singly and doubly occupied sites
$p_e$, $p_s$, $p_d$ can be written as
\bmath
\beq
p_e(n)=(1-\frac{n}{2})^2-(\frac{n}{2})^2\alpha
\eeq
\beq
p_s(n)=n(1-\frac{n}{2})+2\alpha (\frac{n}{2})^2
\eeq
\beq
p_d(n)=(\frac{n}{2})^2(1-\alpha)
\eeq
\emath
for $n\leq 1$, and
\bmath
\beq
p_e(n)=p_d(2-n)
\eeq
\beq
p_s(n)=p_s(2-n)
\eeq
\beq
p_d(n)=p_e(2-n)
\eeq
\emath
for $n> 1$. Here, $\alpha$ is a parameter that describes the 
suppression of double occupancy due to the on-site repulsion $U$,
which we assume to be approximately $n$-independent and to interpolate
smoothly between the limits $\alpha(U=0)=0$ and $\alpha(U=\infty)=1$.
As discussed earlier, the magnitude of $U$ depends on the ionic charge $Z$.
The relative probabilities of singly and doubly occupied sites are then
\bmath
\beq
P_s(n)=\frac{p_s(n)}{p_s(n)+p_d(n)}
\eeq
\beq
P_d(n)=1-P_s(n)
\eeq
\emath
and the incoherent part of the spectral function in Eq. (46) is given by
\beq
A'(\omega)=(1-z(n))[P_s(n) A'_h(\omega)+P_d(n)A_h(\omega)] .
\eeq

In Figure (10) we show the qualitative behavior of the spectral function
Eq. (46) versus band filling for one set of parameters. As the 
band filling starts increasing from zero, the quasiparticle peak starts
decreasing and a peak at positive frequencies forms, corresponding
to the excited states of the doubly occupied ion upon creation of
electrons in singly occupied ions. Only at higher fillings, and
particularly beyond half-filling, where the number of doubly occupied atoms
becomes appreciable, does the peak at negative frequencies start to form,
corresponding to the excited states of the singly-occupied ion
resulting from destruction of an electron in the correlated ground state
of the two-electron atom. The peak at positive energies decreases as
the full band is approached, because we are not considering the possibility
of creation of electrons in higher energy states of the doubly occupied atom.
The quasiparticle peak height decreases monotonically as the band filling
increases, approaching $S_1^2$ as the band becomes completely full.
In the limit $\alpha \rightarrow 1$, corresponding to very strong intra-atomic
Coulomb repulsion (i.e. very large ionic charge $Z$) the peak at 
negative energies only starts to form for $n>1$. Note also, as discussed in
the previous section, that the relative magnitude of the quasiparticle and
the incoherent parts will depend on the ionic charge $Z$, with the
incoherent contributions becoming more important for small $Z$. Furthermore,
the energy scale of the incoherent excitations decreses with decreasing $Z$.

\subsection{Frequency-dependent conductivity}

On general grounds we expect that the frequency-dependent conductivity will
be given by a $\delta$-function at zero frequency broadened to a Drude 
form by disorder and electron-phonon interactions if the spectral function is
of the $\delta$-function form Eq. (31a), i.e. for the case of electrons
in an almost empty band. On the other hand, as the band becomes full and the
single electron spectral function develops incoherent contributions we
expect the frequency-dependent conductivity to develop high frequency
contributions reflecting incoherent absorption processes where the atoms $i$
and/or $j$ end up in excited states when an electron absorbs a photon and jumps from
atom $i$ to atom $j$. A convenient form for the frequency-dependent
conductivity that reflects this physics is given by the form derived
within dynamical mean field theory\cite{dynmf}
\beqn
\sigma_1(\omega)&=&\pi e^2\int_{-\infty}^\infty d\epsilon
\int_{-\infty}^\infty d\omega ' \rho(\epsilon) A(\epsilon,\omega')
A(\epsilon,\omega+\omega ') \\ \nonumber & & \times\frac{f(\omega)-f(\omega+\omega ')}{w} .
\eeqn
Here, $f(\omega)$ is the Fermi function and $A(\epsilon_k,\omega)$ is the
single electron spectral function. $\rho(\epsilon)$ is the 
electronic density of states for the quasiparticle band. Eq. (52) is
exact in the limit of infinite dimensions, and is believed to be a reasonable approximation 
for finite dimensions.

We evaluate Eq. (52) for the phenomenological spectral function given by Eqs. (46) and (51). 
 For the quasiparticle band density of states we choose
for simplicity a triangular form
\beqn
\rho(\epsilon)&=&\frac{2}{D}(1+\frac{\epsilon}{D/2}) ; -\frac{D}{2}<\epsilon<0 \\ \nonumber
&=& \frac{2}{D} (1-\frac{\epsilon}{D/2}) ; 0<\epsilon<\frac{D}{2}
\eeqn
with bandwidth $D=z(n)D_0$, and $D_0$ the 'bare' bandwidth for electrons in the
almost empty band. The 'effective bandwidth' $D$ becomes progressively smaller
as the band filling increases, corresponding to the progressive reduction in
the quasiparticle hopping amplitude
\beq
t(n)=t_0 z(n)
\eeq
with $t_0$ the hopping for the almost empty band. The Fermi energy is related to
the band occupation by
\beqn
\epsilon_F&=&\frac{D}{2} (-1+\sqrt{n}) ; n\leq 1 \\ \nonumber
&=& \frac{D}{2} (1-\sqrt{2-n}) ; n>1
\eeqn
in this model. The frequency-dependent conductivity has three contributions
\beq
\sigma_1(\omega)=\sigma_1^1(\omega)+\sigma_1^2(\omega)+\sigma_1^3(\omega) .
\eeq
At zero temperature, they are given by
\bmath
\beq
\sigma_1^1(\omega)=\pi e^2 z^2(n) \rho(\epsilon_F) \frac{1}{\pi}\frac{\Gamma}
{\omega^2+\Gamma^2}
\eeq
\beqn
\sigma_1^2(\omega)&=&\frac{\pi e^2 z(n)}{\omega}\int_{-\omega}^0
[A'(\omega+\omega')\rho(\epsilon_F+\omega') \\ \nonumber & &+A'(\omega ')\rho(\epsilon_F
+\omega +\omega')]
\eeqn
\beq
\sigma_1^3(\omega)=\frac{\pi e^2}{\omega} \int_{-\omega}^0 d\omega'
A'(\omega ') A'(\omega + \omega ') .
\eeq
\emath
In Eq. (57a), we have replaced the $\omega=0$ $\delta$-function by a Drude-like
lorentzian with relaxation time $\tau=1/\Gamma$ as would be expected in the
presence of impurities or electron-phonon scattering.

The physical content of the different contributions to $\sigma_1(\omega)$ is
as follows. $\sigma_1^1$ describes the intra-band optical absorption
of the quasiparticles, where the atoms remain in their ground state when
the electron hops, and in particular gives the zero frequency conductivity.
Note that it is proportional to $z(n)=1/m^*$ as expected in the Drude 
model (one of the powers of $z(n)$ is cancelled by the density of states
factor). $\sigma_1^2$ has two contributions; the first term has
a non-zero contribution from the incoherent part of the spectral function
at positive frequencies: it corresponds to processes where an electron
hops to a singly occupied site, and the resulting doubly occupied atom
ends up in an excited state, while the atom where the electron hopped
from (initially doubly or singly-occupied) remains in the ground state;
similarly, the second term in $\sigma_1^2$ has a non-zero contribution
from the incoherent part of the spectral function at negative frequencies, 
corresponding to processes where an electron hops from a doubly occupied
site and the resulting singly occupied atom ends up in an excited
state, while the atom the electron hops to (either singly or unoccupied
initially) ends up in the ground state. Finally, $\sigma_1^3$ describes 
processes where an electron hops from a doubly occupied site to a
singly occupied site and the atoms at both sites end up in excited final states.

Figure 11 shows the obtained frequency-dependent conductivity for one set of 
parameters using the phenomenological spectral function discussed in the
previous subsection. For low band filling, with the spectral function dominated
by the quasiparticle part, the optical conductivity is largely given by the
intra-band Drude part, and the zero frequency conductivity is high. In other words,
the system is a 'good metal', with small resistivity. As the
band filling increases, the intra-band conductivity first increases as the
number of electrons increases, however at the same time some of the
increased absorption occurs at higher frequency due to the incoherent part
of the spectral function. At even higher band filling where the intra-band
conductivity becomes hole-like, the Drude part becomes much smaller as does
the zero-frequency conductivity; at the same time, a large incoherent
contribution to the conductivity develops; the spectral weight of the incoherent
part arises from the reduced intra-band contribution due to the reduced
quasiparticle weight. The system is now a 'bad metal', with high resistivity
and largely incoherent optical absorption.

It should be pointed out that our calculated conductivity certainly does not
satisfy the 'global' conductivity sum rule\cite{mahan}
\beq
\int_0^\infty d\omega \sigma_1(\omega)=\frac{\pi e^2 n}{2m}
\eeq
which states that the total conductivity spectral weight should
increase with the number of electrons in the system ($m$ is the bare electron mass). 
This is because our model does not take into account interband transitions,
and because it does not allow for absorption processes where the atomic
final state contains three electrons. Nevertheless, the model correctly
describes the transfer of optical spectral weight from intraband to
higher frequency incoherent absorption as the bandfilling progressively increases.

\subsection{Dc conduction}
As a consequence of this physics, it is clear that the process of electrical conduction 
in a hole metal is intrinsically more complicated than that in an electron metal. When
a hole hops from a site to a neighboring site, there is a finite probability 
amplitude for the atoms to end up in excited states rather than the ground 
state. That is what gives rise to the reduced dc conductivity as expressed
by the enhanced effective mass in the Drude formula. Thus the conduction cannot
be described as an entirely coherent process, as in the case of electrons, where
the phase of the wavefunction is well defined over a mean free path. In the case
of the hole, only the part of the conduction associated with the quasiparticle
part of the spectral function can be described as coherent. When the
quasiparticle weight is small, the transport of current is largely incoherent.

It is of course true that in a perfect translationally invariant crystal the
many-body eigenstates are also eigenstates of the crystal momentum operator\cite{sham}, 
and one
might argue that such a state with finite $\vec{k}$ could carry current without
dissipation. However, any imperfections in the crystal would make such arguments
invalid, unless the system goes into the superconducting state and acquires rigidity.

One may expect new physical phenomena associated with this incoherent transport. 
If an atom ends up in an excited state after the hole hops, it amounts to
creating an electronic excitation in the system, just as the electron-phonon
interaction can give rise to creation of a phonon . In the electron-phonon case
this phonon emission gives rise to Joule heating that dissipates part of the energy 
supplied to
the system by the external potential. Similarly, here the electronic excitation 
should decay through $photon$ emission, with the photon frequency determined
by the electronic excitation energies involved. Because of conservation of energy
it is clear that such photon emission cannot occur in every single hopping
process for moderate voltages applied. However, one would expect
emission of non-thermal high frequency radiation at
locations in the solid where translational symmetry is broken, i.e. impurities and
boundaries.  In particular, electroluminescence\cite{elum} (at visible and infrared
frequencies) would be expected to
occur in this picture at the sample-$positive$ electrode boundary, where
holes are injected into the hole metal sample. The effect should be strongest in samples
where the holes are highly dressed in the normal state, i.e.  
where the band is close to full and the ionic charge $Z$ is small. Such systems
should also give rise to high $T_c$ superconductivity, as discussed in the
next section. When the system
goes superconducting transport becomes coherent and the photon emission
should dissappear.

Furthermore, the physics of hole conduction discussed here suggests that the
contact resistance between an electron-metal electrode and a highly dressed
hole metal sample should be non-symmetric. When an electron is injected into the
hole metal it needs to excite the other electron in the orbital, while
when an electron is removed from the  hole metal the remaining electron
relaxes to a lower energy state. Hence electron injection into the hole
metal should offer
more resistance than electron removal, giving rise to a rectifying
contact where the resistivity is lower at the positive electrode-hole metal
contact.

\section{Hole superconductivity}

The physics discussed earlier leads to a mechanism for superconductivity\cite{hole1,hole2,undr}.
Consider the low energy effective Hamiltonian for quasiparticles Eq. (44):
the hopping amplitude for a quasiparticle depends on the occupation of the sites
involved in the hopping process. For zero, one and two other electrons
in these sites the hopping amplitude for an electron is respectively
\bmath
\beq
t_{ij}^0=t_{ij}
\eeq
\beq
t_{ij}^1=S_1 t_{ij}
\eeq
\beq
t_{ij}^2=S_1^2t_{ij} \equiv \tilde{t}_{ij}
\eeq
\emath
becoming progressively smaller as the occupation increases, due to the
progressive reduction in quasiparticle weigth. It is convenient to rewrite
the quasiparticle kinetic energy in terms of hole rather than electron
operators through a particle-hole transformation (that doesn't change the
physics). The quasi-hole Hamiltonian is then
\bmath
\beq
H_{qh}=-\sum_{i,j,\sigma}[\tilde{t}_{ij}+\Delta t_{ij}
(\tilde{n}_{i,-\sigma}+\tilde{n}_{j,-\sigma})]\tilde{c}_{i\sigma}^\dagger
\tilde{c}_{j\sigma} +\sum_{i,j}U_{ij}\tilde{n}_i\tilde{n}_j
\eeq
\beq
\Delta t_{ij}=t_{ij}^1-t_{ij}^2=
(\frac{1}{S_1}-1)\tilde{t}_{ij}\equiv \Upsilon \tilde{t}_{ij} .
\eeq
\emath
Here, the operators are hole rather than electron operators. We have also added
a term describing density-density Coulomb repulsion between holes $U_{ij}$, of
which the largest will be the on-site repulsion $U\equiv U_{ii}$. Furthermore, we
have omitted the term describing hopping of a hole in the presence of two other
holes, which should be negligible for low hole density due to Coulomb
repulsion.

Superconductivity will occur for this model Hamiltonian  in the regime
of low hole density (band almost full) for a range of parameters, driven
by lowering of kinetic energy when hole carriers pair\cite{hole1,hole2}. For simplicity we
now restrict ourselves to the case of nearest neighbor hopping and
interactions only. The hole quasiparticle Hamiltonian is
\beqn
H&=&-\sum_{<ij>,\sigma} [t+\Delta t(n_{i,-\sigma}+n_{j,-\sigma})]
(c_{i\sigma}^\dagger c_{j\sigma}+h.c.) \nonumber \\
& &+U\sum_i n_{i\uparrow} n_{i\downarrow}
+V\sum_{<ij>}n_i n_j .
\eeqn
We have omitted the 
'tilde' on the hole quasiparticle operators. This 'correlated hopping'
Hamiltonian has been extensively studied in recent years by approximate
and exact techniques, and the reader is referred to the references for
detailed information\cite{hole1,h0,h1,h2,h3,h4,h5,h6,h7,h8,dilute}. 
The condition for superconductivity in the limit
of low hole concentration is\cite{dilute}
\beq
K>\sqrt{(1+u)(1+w)}-1
\eeq
with $K=2z\Delta t/D$, $u=U/D$, $w=zV/D$, with $z$ the number of nearest
neighbors to a site and $D$ the (renormalized) bandwidth. Within BCS
theory the same condition is found, with $1/D$ replaced by $g(\epsilon_F)$, the
density of states at the Fermi energy, in the definition of the parameters
in Eq. (62)\cite{hole1}. The reader is referred to the literature for discussion of the particular
features of the superconducting state (e.g. energy dependent gap function) and
for results for various observables (e.g. density and pressure dependence of
$T_c$, thermodynamics, tunneling, NMR, disorder effects, etc.) that can be
understood from this low energy effective Hamiltonian\cite{hole1,hole2}.

Here we wish to focus the discussion on the fundamental aspect of the physics
related to the high energy degrees of freedom that are not contained in the low 
energy quasiparticle Hamiltonian\cite{color,undr}. The key to superconductivity 
lies in the relation between
the relative weights of the coherent and incoherent parts of the spectral function
discussed in Sect. III and the site occupation, or the band filling.
As we have seen, as the electronic site occupation increases the spectral
function becomes more incoherent. Consider the Cooper pair wavefunction
for a hole pair in the full band:
\beq
\Psi=\frac{1}{\sqrt{N}} \sum_{i,j} f_{ij} 
c_{i\uparrow}^\dagger c_{i\downarrow}^\dagger |0>
\eeq
where $|0>$ is the full band, and $c_{i\sigma}^\dagger$ is a hole
creation operator. In the limit where the size of the Cooper pair
(i.e. the superconducting coherence length) goes to infinity the holes
become uncorrelated and $f_{ij}\sim 1/\sqrt{N}$ ($N=$number of sites in
the system). For short coherence length instead, $f_{ij}\sim O(1)$ for
$|i-j|$ small. In particular, the amplitude for the holes to be
on the same site or on neighboring sites becomes of order unity rather
than being negligible. Thus, the spectral function for a hole in such
a state will have a larger coherent part and a smaller incoherent part
than the spectral function of a hole that is not bound in a pair. 
In other words, holes 'undress', i.e. become more coherent,  when they pair\cite{undr}.
Because electrons (=quasiparticles when the band is almost empty) are
more coherent than holes (=quasiparticles when the band is almost full)
we may also say that holes become more 'like electrons' when they
bind in a Cooper pair. Naturally, if 
pairing and superconductivity is associated with such quasiparticle 'undressing', 
it will occur when the quasiparticles in the normal state are hole-like (band close to full)
and will not occur when they are electron-like (band close to empty)\cite{chapnik}. 

Intimately related to the increased coherence of the hole bound in a Cooper pair is the fact
that the hole has an increased hopping amplitude when it hops to or from a
site where there is another hole, due to its increased quasiparticle weight. 
When a hole forms a Cooper pair the probability for another hole to be
nearby increases, and the increased
hopping amplitude that occurs on the average 
gives rise to a lowering of the
hole kinetic energy upon pairing, which provides the binding energy of the
Cooper pair if it can overcome  the increased cost in ordinary Coulomb energy arising
from the decreased distance between like charges\cite{apparent}.

Finally, consider the process of optical absorption when the band is almost full. 
An incident  photon causes
an electron to hop from a site to a neighboring site. If the electron is
part of a hole Cooper pair, it is far more likely to hop to or from a site
that is $unoccupied$ by another electron, than if it is uncorrelated. If the electron
hops to or from an unoccupied site, it has  a larger contribution from the
coherent part of the spectral function and hence gives a larger contribution to
the 'intraband' part of $\sigma_1$ rather than to the high frequency
incoherent part. Hence, optical spectral weight will shift
from the high frequency incoherent part of the spectrum
to the  low frequency Drude part when pairing occurs. Because in the
superconducting state the lowest frequency part of the Drude absorption
(below twice the energy gap $\Delta$) collapses into the zero frequency
$\delta-$function that determines the penetration depth, the
$\delta-$function acquires an 'anomalous' extra contribution 
transferred from the high energy incoherent optical absorption spectrum rather than
from the frequency region below $2\Delta$, and an 'apparent violation'
of the conductivity sum rule results\cite{apparent}.

In summary, the superconducting transition driven by the physics of 
electron-hole asymmetry discussed here has associated with it these
anomalous spectral weight transfers in one- and two-particle (optical
conductivity) spectral functions, caused by the dependence of the single
particle spectral function on site occupation discussed in Sect. III, 
$together$ $with$ $the$ $fact$
$that$ $the$ $site$ $occupation$ $changes$ $when$ $pairing$ $occurs$. The spectral weight
is transfered from an energy range that is unrelated to the magnitude of
the superconducting energy gap $\Delta$, rather it is related to the much larger
energy scale of intra-atomic electronic excitations. To study 
quantitatively these spectral weight transfer processes it is useful to
consider model Hamiltonians as discussed in the next section.

\section{Model Hamiltonians}

To go beyond the phenomenological model discussed in the previous section
it is useful to construct model Hamiltonians that contain the physics of
interest, that are amenable to theoretical studies by various techniques.
We consider here three classes of model Hamiltonians.

\subsection{Hamiltonians with auxiliary oscillator degrees of freedom}

A variety of generalizations of the commonly used Holstein model describing
coupling of a local boson (usually phonon) degree of freedom to the local
electronic charge density can be constructed to contain the essential physics
discussed here\cite{polaron}. As the simplest example, consider an electronic
model with a single orbital per site and a local boson degree of freedom,
with site Hamiltonian
\beq
H_i=\frac{p_i^2}{2M}+\frac{1}{2} K q_i^2 +(U+\alpha q_i)
n_{i\uparrow}n_{i\downarrow}
\eeq
describing the coupling of a local oscillator $q_i$ of frequency $\omega_0=\sqrt{K/M}$
to the atomic $double$ $occupancy$. Completing the squares yields
\bmath
\beq
H_i=\frac{p_i^2}{2M}+\frac{1}{2}K(q_i+\frac{\alpha}{K}n_{i\uparrow}n_{i\downarrow})^2
+U_{eff}n_{i\uparrow}n_{i\downarrow}
\eeq
\beq
U_{eff}=U-\frac{\alpha^2}{2K}\equiv U-\omega_0 g^2
\eeq
\emath
so the equilibrium position of the oscillator is $q=0$ for the empty and
singly occupied site, and $q=-\alpha/K$ for the doubly occupied site. The reduction 
from $U$ to $U_{eff}$ due to the change in the oscillator equilibrium position
parallels the reduction from $U$ to $U_{eff}$ in the atomic problem due
to the modification of the electronic wave function in the doubly occupied
atom by correlations.

The states of the site can be written as direct product of the electronic state
in occupation number representation and the boson state. We denote by 
$|n^l>$ the $l-$th excited state of the oscillator when there are $n$ electrons
at the site (n=0, 1 or 2) , and $|n^0>\equiv |n>$ the ground state.
The site ground states are then $|0>|0>, |\uparrow>|1>, |\downarrow>|1>,
 |\uparrow \downarrow>|2>$. Because the oscillator is not affected by
single occupancy, $|0^l>=|1^l>$. The matrix elements $S_l$ defined in Eq. (13c) and
excitation energies Eq. (13b) are given by
\bmath
\beq
S_{l+1}=<1^l|2>=\frac{e^{-g^2/2} g^l}{(l!)^{1/2}}
\eeq
\beq
E_l^1=\omega _0 l
\eeq
\emath
and the spectral function for hole creation  in the doubly occupied
site $A_h(\omega)$ has the form given by Eq. (14), with quasiparticle weight
\beq
z=S_1^2=e^{-g^2} .
\eeq
The spectral functions for electron creation in the empty site and electron destruction
in the singly occupied site $A_{el}$ and $A'_{el}$ are single $\delta-$functions, 
as in the case of the atom, because the oscillator
is not affected by single occupancy.  Finally, the spectral function for hole 
destruction (electron creation) in the singly occupied site $A'_h(\omega)$ is
a sum of $\delta-$functions like Eq. (21), with
\bmath
\beq
S'_l=(-1)^l S_l
\eeq
\beq
E_l^2=E_l^1
\eeq
\emath

In an atom, the energy level spacing is proportional to $Z^2$, with $Z$ the nuclear
charge. If we assume similarly
\bmath
\beq
\omega_0=cZ^2\eeq
and we also assume
\beq
g^2=\frac{c'}{Z^2}
\eeq
\emath
(c, c' constants) then the reduction from $U$ to $U_{eff}$ in Eq. (65b) is 
independent of $Z$, which is what is found in real atoms\cite{sla}. Furthermore, as
$Z$ decreases $g$ increases and the quasiparticle weight Eq. (67) will decrease, similarly
as found in the atomic problem in Sect. III. For the solid we take then
as model Hamiltonian
\beq
H=-\sum_{ij} t_{ij} c_{i\sigma}^\dagger c_{j\sigma} +\sum_i H_i
\eeq
and argue that this model Hamiltonian describes the same physics as the real
solid discussed earlier. In particular, the low energy effective 
Hamiltonian is of the same form as Eq. (44), and the same processes of
spectral weight transfer discussed earlier will occur here. Quasiparticle
operators are easily defined using a generalized Lang-Firsov transformation\cite{undr,color}.

This Hamiltonian can be studied by standard theoretical techniques. Some initial
studies have already been performed\cite{undr,dynhub}. Further studies with non-perturbative
techniques such as dynamical mean field theory\cite{dynmf,dynmf2} and numerical methods such as
density-matrix renormalization group\cite{dmrg} should be able to determine quantitatively
the parameter regime where superconductivity occurs in this model, as well
as provide substantial insights
into the physics of electron-hole asymmetry and spectral weight transfer
under discussion here.

It should be pointed out however that there is a difference between this
Hamiltonian and the atomic case previously discussed. Namely, both the 
excitation spectrum and matrix elements here are the same for hole destruction
in the doubly occupied site and electron creation in the singly occupied site,
unlike the case of the real atom. However, we do not believe that this
difference will qualitatively change the physics.

It is also possible to extend the Hamiltonian Eq. (64) so that it will also
describe dressing of electrons in singly occupied sites\cite{polaron}, by including coupling
to the electronic charge density as in the ordinary Holstein model. As long as 
the coupling to double occupancy is non-zero, or through other terms that
break electron-hole symmetry such as anharmonicity\cite{polaron}, the essential
physics of interest here, that as the band filling increases the coupling strength and
hence the dressing of the quasiparticle increases, will be contained in the Hamiltonian.  
Such a generalized Holstein model would be useful if one
is interested in describing the smaller dressing effects of the first electron in
a shell due to coupling to electrons in inner shells.

\subsection{Electronic Hamiltonian}

An electronic Hamiltonian without auxiliary degrees of freedom that
contains the physics of electron-hole asymmetry and undressing under
consideration here needs to have at least two orbitals per site. A simple such
Hamiltonian for the site is\cite{electron}
\beqn
H_i&=&Un_{i\uparrow}n_{i\downarrow}+U'n'_{i\uparrow}n'_{i\downarrow}+
V n_i n'_i +\epsilon n'_i \nonumber \\
& & -t'\sum_\sigma (c_{i\sigma}^\dagger c'_{i\sigma}+h.c.)
\eeqn
where the primed and unprimed orbitals refer to electrons in the 
upper (lower) energy orbital, assumed orthogonal to each other. (An alternative
description would assume $t'=0$ and non-orthogonal orbitals). We assume all
parameters in Eq. (71) positive, and the ordering
\bmath
\beq
U'+2\epsilon < V+\epsilon <U
\eeq
\beq
U,U',V >> \epsilon >>t'   .
\eeq
\emath
These conditions ensure that a single electron resides primarily in the lower
level and two electrons reside primarily in the upper level. Hence this
Hamiltonian mimics the physics of the Hartree Fock solution of the atomic 
problem discussed previously. The effective on-site repulsion defined 
by Eq. (7) is $U_{eff}=U'+2\epsilon$ to lowest order in $t'$.

It is simple to diagonalize the site Hamiltonian exactly, however for
illustration purposes we restrict ourselves here to lowest order
perturbation theory in $t'$. The ground states of the singly and
doubly occupied sites, $|\sigma>>$ and $|\uparrow \downarrow>>$ and their
energies are, respectively
\bmath
\beq
|\sigma>>=|\sigma>|0>+\delta_1 |0>|\sigma>  ; E=0
\eeq
\beq
|\uparrow \downarrow>>=|0>|\uparrow\downarrow>+\delta_2\frac
{|\uparrow>|\downarrow>+|\downarrow>|\uparrow>}{\sqrt{2}} ; E=U'+2\epsilon
\eeq
\emath
where the first ket refers to the lower and the second ket to the upper 
orbital, and 
\bmath \beq
\delta_1=\frac{t'}{\epsilon}
\eeq
\beq
\delta_2=\frac{\sqrt{2} t'}{V-U'-\epsilon}
\eeq \emath
The excited state of the singly occupied atom is
\beq
|\sigma^{(2)}>>=\delta_1 |\sigma>|0>-|0>|\sigma>  ; E=\epsilon
\eeq
and the excited states of the doubly occupied atom are
\bmath \beqn
|\uparrow\downarrow^{(2)}>>&=&\frac{|\uparrow>|\downarrow>+|\downarrow>|\uparrow>}{\sqrt{2}}
+\delta_3|\uparrow\downarrow>|0> \\ \nonumber
& &-\delta_2|0>|\uparrow\downarrow> ; E=V+\epsilon
\eeqn
\beq
|\uparrow\downarrow^{(3)}>>={|\uparrow>|\downarrow>-|\downarrow>|\uparrow>}{\sqrt{2}}
 ; E=V+\epsilon
\eeq
\beq
|\uparrow\downarrow^{(4)}>>=|\uparrow\downarrow>|0>-\delta_3
{|\uparrow>|\downarrow>+|\downarrow>|\uparrow>}{\sqrt{2}}  ; E=U
\eeq
\emath
with $\delta_3=\sqrt{2} t'/(U-V-\epsilon)$.
The spectral functions to create an electron in the empty site and
to destroy an electron in the singly occupied site are single $\delta-$functions, as in the
atomic case. The 'hole wavefunction' $\varphi_\alpha$ that maximizes the hole
quasiparticle weight is given, to this order in $t'$, simply by the
wavefunction of the upper state. The spectral function for hole creation in the
doubly occupied atom is then of the general form Eq. (14):
\bmath \beq
A_h(\omega)=z\delta(\omega)+S_2^2\delta(\omega+\epsilon)
\eeq
\beq
z=S_1^2=(\delta_1+\frac{\delta_2}{\sqrt{2}})^2
\eeq
\beq
S_2=1-\frac{\delta_1\delta_2}{\sqrt{2}}
\eeq \emath
and the spectral function for destroying a hole in the singly occupied
atom, i.e. creating an electron, is
\bmath \beqn
A'_h(\omega)&=&z\delta(\omega)+(S'_2)^2\delta(\omega-(V-U'-\epsilon)) \nonumber \\
& &+
(S'_3)^2 \delta(  \omega-(U-U'-2\epsilon))
\eeqn
\beq S'_2= \frac{1}{\sqrt{2}}-\delta_1\delta_2
\eeq \beq S'_3=\frac{1}{\sqrt{2}} \eeq \beq S'_4=\delta_3 \eeq
\emath
Hence the site spectral functions are of the same form as in the atomic
problem discussed previously. The 'quasiparticle operators' are then
$c_{i\sigma}$ or $c'_{i\sigma}$ depending on whether the site is
singly or doubly occupied. We may take as lattice Hamiltonian
\beqn
H&=&-\sum_{ij}t_{ij}[c_{i\sigma }^\dagger c_{j\sigma }+
c_{i\sigma }^\dagger c'_{j\sigma }+(c'_{i\sigma })^\dagger c_{j\sigma }+
(c'_{i\sigma })^\dagger c'_{j\sigma }]\nonumber \\
& &+ \sum_i H_i
\eeqn
where we have assumed for simplicity the same hopping amplitude for the
unprimed and primed orbitals. Projection of the Hamiltonian Eq. (79) to the
lowest site energy states yields a low energy effective Hamiltonian of
the form Eq. (44), which will give rise to superconductivity in certain
parameter ranges. Furthermore the spectral weight transfer processes
discussed previously should occur for this Hamiltonian. Because it has
a relatively small number of states per site this Hamiltonian should 
be amenable to study particularly by exact diagonalization methods.
Parameters in this Hamiltonian can be related fairly directly to 
realistic parameters obtained from quantum chemical calculations
of real atoms.

\subsection{Hamiltonians with auxiliary spin degrees of freedom}

The physics of interest here was first studied in the context of an
electronic Hamiltonian with an auxiliary spin 1/2 degree of freedom\cite{spin,spin2}
to represent the other electron(s) in the atom, with
site Hamiltonian
\beq
H_i=[V(n_{i\uparrow}+n_{i\downarrow})-\omega_0]\sigma_z^i + \Delta \sigma_x^i +
U n_{i\uparrow}n_{i\downarrow}
\eeq
with $V>\omega_0$ and $\Delta << \omega_0 , V$, and all parameters positive.
For such parameters, the auxiliary spin  in the ground state will point predominantly
up if $n_{i\uparrow}=n_{i\downarrow}=0$ and down otherwise. That situation
describes then a $hole$ Hamiltonian, where the first hole causes a large
disruption in the 'background' degree of freedom (the spin)
and the second a small one. We can however also use this 
Hamiltonian with the operators representing $electrons$ if we choose the parameter
range $V<\omega_0<2V$ instead, so that the first electron will not change
the state of the spin and the second one will. The lattice Hamiltonian
\beq
H=-\sum_{ij} t_{ij} c_{i\sigma}^\dagger c_{j\sigma}+\sum_i H_i
\eeq
also has as low energy effective Hamiltonian the generic form Eq. (44) that gives rise
to correlated hopping\cite{hm}. Exact diagonalization studies of this Hamiltonian on
small clusters\cite{spin2} show that it gives rise to pairing in a wide range of 
parameters. Furthermore, the optical conductivity was calculated
in a simple case\cite{color} and the results show the spectral weight
transfer process discussed earlier, as one would expect.

The site spectral functions are also simple to calculate. There is a 
difference with the atomic case discussed previously in that the single
electron spectral function for the empty site is not a single $\delta-$function
here because some modification of the background spin degree of freedom
occurs even when the first electron is created at the site. Nevertheless, for
the parameter range discussed the change in the state of the background is
much larger for the second electron, hence the quasiparticle weight is 
much smaller for a hole than for an electron and the same qualitative 
physics of electron-hole asymmetry and undressing occurs. Here again,
this Hamiltonian should be amenable to detailed study by exact diagonalization.

Finally, a site Hamiltonian analogous to the oscillator Hamiltonian Eq. (64) can 
also be constructed with auxiliary spin degrees of freedom:
\beq
H_i=[Vn_{i\uparrow}n_{i\downarrow}-\omega_0]\sigma_z^i + \Delta \sigma_x^i +
U n_{i\uparrow}n_{i\downarrow}
\eeq
which will contain the physics of interest here for $V>\omega_0$. This 
Hamiltonian does not 'dress' the single electron as Eq. (80) does.
A two-parameter version that resembles closely the oscillator
Hamiltonian would be
\beq
H_i=\omega_0\sigma_z^i+g\omega_0[1-2n_{i\uparrow}n_{i\downarrow}]\sigma_x^i
+U n_{i\uparrow}n_{i\downarrow}
\eeq
which is analogous to the oscillator Hamiltonian Eq.  (64) which in
terms of boson creation and annihilation operators reads
\beq
H_i=\omega_0a_i^\dagger a_i+(U+g\omega_0(a_i^\dagger +a_i))n_{i\uparrow}n_{i\downarrow} .
\eeq
For large $g$, the Hamiltonians Eq. (83) and (84) give rise to small
quasiparticle weight for the hole.

In summary, there is a large variety of model Hamiltonians that can be
constructed that contain the essential physics of interest here,
and we discussed some particular simple examples. Study of these
Hamiltonians should yield interesting insights and clarify the
physics further. However this physics will only exist for certain parameters
in the models, and not for others. In particular, the models need to
break electron-hole symmetry. Instead, if one takes for example the
spin Hamiltonian Eq. (80) with $V=\omega_0$, or the Holstein Hamiltonian
with only coupling to the local charge density, the physics of interest
is lost because the models in those cases are electron-hole symmetric.

\section{Discussion}

In this paper we have addressed the microscopic reasons for why holes are not 
like electrons in condensed matter. We have shown that there are qualitative
differences between electrons and holes in atoms that have fundamental
consequences for the physics of the solid state. We have proposed that
the essential physical difference relates to the relation between the coherent
(quasiparticle part) and incoherent parts of the one-electron spectral function,
and the electron versus hole character of the quasiparticle. Namely, we have
argued that, reflecting the physics at the atomic level, as the filling of a band 
increases the quasiparticle becomes
progressively less coherent while at the same time the character of the
quasiparticle changes from electron-like to hole-like. Furthermore, 
the effective mass of the quasiparticle is inversely proportional to the degree
of coherence (quasiparticle weight) and hence increases as the band filling increases.

This physics naturally leads to a mechanism of superconductivity, since 1) superconductivity
involves pairing of quasiparticles, and 2) when hole quasiparticles
 pair, the $local$ $density$
$of$ $electrons$ $decreases$. Hence, as holes bind in a Cooper pair they become more
coherent, and more mobile, and more like electrons, and the superconducting
state is stabilized by the accompanying lowering of kinetic energy. As a result
the superconducting state is more coherent than the normal state, and
spectral weight at low energies in one- and two-particle spectral functions
increases while incoherent spectral weight at high energies decreases.
In other words, quasiparticles 'undress' when they pair\cite{undr}.

The informed reader may wonder about the relation between the parameters in the
low energy effective Hamiltonian Eq. (44) and the ones calculated in our
earlier work on hopping amplitudes in the diatomic molecule\cite{molec}.
For example, Eq, (44) implies that $t_2/t_1=t_1/t_0$, with $t_i$ the 
hopping amplitude for an electron when there are $i$ other electrons at
the 2 sites involved in the hopping process. In fact, the results found
in Ref. \cite{molec} do not generally obey this relation. The answer is, the
calculation in Ref. \cite{molec} is certainly expected to be more accurate.
However, the derivation here reflects the essential physics in a clearer way
and allows to discuss the fundamental processes of spectral weight transfer. The
two calculations are of course intimately related, in particular recall that
in the calculations in ref. \cite{molec} the difference in hopping
amplitudes went to zero as the atomic charge became large or if one 
artificially constrained the atomic orbitals to not change with occupation,
in agreement with the physics discussed here.

One may argue that the difference between electrons and holes discussed here
is strictly speaking only a $qualitative$ difference for the lowest $1s$ band. For higher
bands, even a single electron in an empty band will be partially 
dressed from interactions with electrons in lower bands (not to mention 
electron-phonon interactions). While this is true, the considerations
in this paper certainly imply that within a given band
\beq
z(n)>z(2-n)
\eeq
for $n\leq 1$, that is, the hole at band filling $(2-n)$ is always more
dressed than its counterpart electron at band filling $n$. Hence, the mechanism
of superconductivity whereby holes in a band pair to 'undress' and become more
like the electrons in that band should still be operative.

There is in fact another qualitative  difference between electrons and holes
in energy bands, i.e. between electrons at the bottom and electrons at the
top of the band. Namely, the wavefunction for the quasiparticle at the bottom
of the band has a large amplitude in the region between the
atoms, thus providing stability to the lattice (bonding state). Instead,
the wavefunction for the quasiparticle at the top of the band  has vanishing
amplitude in the region between the atoms, as it has highest energy and is orthogonal 
to all the band states below it (antibonding state). Correspondingly, the amplitude of the 
wavefunction at the atoms is larger for electrons at the top  than for electrons
at the bottom of the band, and as a consequence the
dressing due to intra-atomic Coulomb repulsion also from electrons in other
atomic orbitals is larger for electrons at the top of the band.

The phase space of model Hamiltonians that can be written down and studied
as possible models for condensed matter phenomena is infinite. What makes
some models more suitable to describe real materials than others? We propose
that a useful criterion to validate one Hamiltonian over another is whether
it can be derived from a more fundamental Hamiltonian that describes physics
on a larger energy scale. In particular: the low energy Hamiltonian Eq. (44)
(or Eq. (61)) describes an $increase$ of low energy spectral weight when
carriers pair\cite{apparent}. By itself, it does not contain the
high energy degrees of freedom where that spectral weight is coming from. However,
it can be derived from a more fundamental Hamiltonian (e.g. the models discussed
in Sect. VII) that does contain the high energy degrees of freedom where the
spectral weight comes from. In contrast, the attractive Hubbard model
for example can also describe pairing of carriers, and it describes an
associated $decrease$ in low energy spectral weight (because the effective 
mass increases upon pairing); however, it is not clear that it is possible
to derive an attractive Hubbard model from a more general Hamiltonian that
would describe a corresponding $increase$ in high energy spectral weight.

As we have discussed, there are a variety of models that one can study that
contain the physics of interest here. Thus, the issue is not the validity
of a given model Hamiltonian, rather it is the validity of a very general
paradigm. Namely, that the conventional description of energy bands in solids
needs to be augmented, beyond the knowledge of the $\epsilon$ versus $k$
relation and the quasiparticle wavefunction, to the description of the
behavior of the coherent and incoherent parts of the one-particle spectral function
as function of band filling, and that these quantities have the particular dependence on
band filling discussed here.

The concepts discussed here are clearly not restricted to one particular class of
solids but rather are generally applicable to all solids, just as the concept of
energy bands is. Thus, it is unreasonable to expect that the mechanism
of superconductivity that results from these concepts would apply to only
one material or any one class of materials, or even that it would apply to
most but not all superconducting materials. In a sense it is like superconductivity
itself: the collective state is robust because it forms a coherent whole, and
cannot be destroyed by local impurities or imperfections in a given solid. 
Similarly the theory discussed here is a seamless web, that cannot be
destroyed by individual examples that seemingly contradict it: it either
applies to all superconductors or to none, and the
apparent counterexamples (e.g. superconductors that seem to have only $electron$ 
carriers in the normal state, or superconductors that exhibit seemingly incontrovertible
evidence for an electron-phonon mechanism) will eventually be explainable
within the same framework, if the framework is valid. So what is the evidence
in favor of this framework?

Most importantly, there is significant empirical evidence that the presence
of hole carriers favors superconductivity\cite{chapnik}. Moreover, the
materials with highest $T_c$ (high $T_c$ cuprates and $MgB_2$) have holes
conducting through negative ion networks ($O^=$ and $B^-$), i.e. small
ionic charge $Z$, which favors superconductivity within this framework.
Moreover, the high $T_c$ cuprates show clear evidence for 'undressing',
both in the normal state as holes are added and at fixed hole concentration
as the system becomes superconducting\cite{undr1,undr2,undr3,undr4,undr5,undr6}, 
as described (in fact $predicted$\cite{hole1,color})
by the present framework. Moreover, the dependence of $T_c$ on doping in 
transition metal intermetallics\cite{matthias} follows the behavior predicted by this
framework\cite{transition1,transition2}. Moreover, the general empirical relation
observed between superconductivity and lattice instabilities follows
naturally from the fact that $antibonding$ states need to be occupied in 
order to have hole carriers. Other experimental evidence in support of
hole superconductivity is discussed
in the references\cite{hole1,hole2,apparent,mgb2}.

There are many open questions within the framework discussed in this paper.
A convenient starting point for the discussion are the model Hamiltonians 
discussed in Sect. VII. First, are there other, similar or otherwise,
model Hamiltonians that should be studied, possibly more appropriate than
those? Next one can move from there in two opposite directions. In
one direction, the study of the properties of these Hamiltonians by
various theoretical techniques. In particular, dynamical mean field 
theory\cite{dynmf2} is likely to be a very fruitful approach.
One should calculate  the most basic physical properties such 
as phase diagram, quasiparticle weight, spectral functions and various
correlation functions. How do the basic superconductivity parameters
such as coherence length, superfluid weight, etc. depend on the 
Hamiltonian parameters?
How do the various spectral weight transfer processes depend on the parameters?
 And in particular, which  observable properties should be calculated
that can differentiate this mechanism from others? In the opposite direction,
one would like to connect the model Hamiltonians to real materials.
What quantities should one evaluate in a first-principles calculation
of a given solid that would be relevant to fix the parameters in the
model Hamiltonians? Once that question is clarified a vast area of
study will open up where any given existing or proposed material could
be tested to see where its parameters lie, in particular concerning
its superconducting properties.

\acknowledgements
The author is grateful to Carlo Piermarocchi, Lu Sham and Harry Suhl for
stimulating discussions.

\appendix
\section{Hylleraas dritte Naherung}

For the convenience of the reader we summarize here the equations that 
determine the parameters in the Hylleraas wavefunction Eq. (5)\cite{hyl}. The energy
(in Ry) is given by $E=Z^2\lambda$, with
\beq
\lambda=-\frac{L^2}{4NM}
\eeq
\bmath
\beq
M=8+50c_1+96c_2+128c_1^2+584c_1c_2+1920c_2^2
\eeq
\beq
N=4+35c_1+48c_2+96c_1^2+308c_1c_2+576c_2^2
\eeq
\beqn
L&=&16-\frac{5}{Z}+(120-\frac{32}{Z})c_1+(192-\frac{36}{Z})c_2
+\\ \nonumber & &(288-\frac{70}{Z})c_1^2 +(1120-\frac{192}{Z})c_1c_2
+(2304-\frac{312}{Z})c_2^2
\eeqn
\emath
Minimization of $\lambda$ yields the coefficients $c_1$ and 
$c_2$ in the wavefunction, and the orbital exponent $k$ is given by
\beq
k=\frac{L}{2M}
\eeq

\section{Calculation of matrix elements $S_n$ in the various approximations}

For the Hartree wavefunction we have for the overlaps Eq. (13c)
\beq
S_n^H=\int d^3r \varphi_{\bar{Z}}\varphi_n(r)=(\varphi_{\bar{Z}},\varphi_n)\equiv
S_n(Z,\bar{Z})
\eeq
These integrals are simply evaluated for the wavefunctions Eq. (26) and yield
\beq
S_n^H=\frac{4(\frac{Z}{n}\bar{Z})^{3/2}}{(\bar{Z}+\frac{Z}{n})^3}
\sum_{l=0}^{n-1}(-1)^l (l+2)! a_l \frac{(\frac{Z}{n})^l}
{(\bar{Z}+\frac{Z}{n})^l}
\eeq
with $a_l$ given by Eq. (26b).

For the Eckart wave function Eq. (4) we have simply
\beq
S_n^E=\frac{S_1(Z_1,\bar{Z})S_n(Z_2,Z)+S_1(Z_2,\bar{Z})S_n(Z_1,Z)}
{[2(1+S_1(Z_1,Z_2)^2)]^{1/2}}
\eeq

For the Hylleraas wavefunction, Eq. (6), we use the relation

\beq
\int d^3r_1 d^3r_2 F=2\pi^2 \int_0^\infty ds \int_0^s du \int_0^u dt u(s^2-t^2)F
\eeq
valid for any symmetric function $F(\vec{r}_1,\vec{r}_2)$, with 
$s$,$t$, $u$ given by Eq. (6). The integrals needed are of the form
\beq
I(p,q,r,a,b)=\int_0^\infty ds \int_0^s du \int_0^u dt u(s^2-t^2)
e^{-as}e^{-bt} s^pt^qu^r
\eeq
and yield
\beqn
&I&(p,q,r,a,b)=\frac{1}{b^{p+q+r+3}} \big{[} \sum_{j=0}^q \frac{q!}{(q-j)!}
\nonumber \\ & &\big{[}\sum_{i=1}^{q+r+1-j}\frac{1}{(\frac{a}{b}+1)^{p+i}}
\frac{(p+i-1)!(q+r+1-j)!}{(i-1)!} \nonumber \\
&-&\frac{p!}{(\frac{a}{b}+1)^{p+1}}(q+r-j)!\big{]}
+\frac{q!(p+r+1)!}{(r+1)(\frac{a}{b})^{p+r+2}} \big{]}
\eeqn
and the overlaps are given by
\beqn
&S&_n=\frac{\pi}{(2Zk)^6} \frac{(\bar{Z}^3(\frac{Z}{n})^3)^{1/2}}{b^6}
\sum_{l=0}^{n-1} (-1)^l\frac{a_l}{b^l}\frac{1}{(4kn)^l}  \nonumber\\
& &\sum_{i=0}^l \frac{l!}{i!(l-i)!}  \Bigl[\bigl[I(2+l-i,i,1,a,b) -\nonumber \\
& & I(l-i,2+i,1,a,b,)+(-1)^{i+l}(I(2+l-i,i,1,a,-b)- \nonumber\\
& &I(l-i,2+i,1,a,-b))\bigr]+\frac{c_1}{b}
\bigl[I(2+l-i,i,2,a,b)-   \nonumber \\
& &I(l-i,2+i,2,a,b,)-
 (-1)^{i+l}(I(2+l-i,i,2,a,-b)- \nonumber\\
& & I(l-i,2+i,2,a,-b))\bigr]+\frac{c_2}{b^2}
\bigl[I(2+l-i,2+i,1,a,b)-\nonumber \\
& & I(l-i,4+i,1,a,b,)+(-1)^{i+l}(I(2+l-i,2+i,1,a,-b)  \nonumber\\
& &-I(l-i,4+i,1,a,-b))\bigr]\Bigr]
\eeqn
with
\bmath
\beq
a=\frac{1+\lambda_1+\lambda_2}{2}
\eeq
\beq
b=\frac{\lambda_2-\lambda_1}{2}
\eeq
\beq
\lambda_1=\frac{\bar{Z}}{2Zk}
\eeq
\beq
\lambda_2=\frac{1}{2kn}
\eeq
\emath

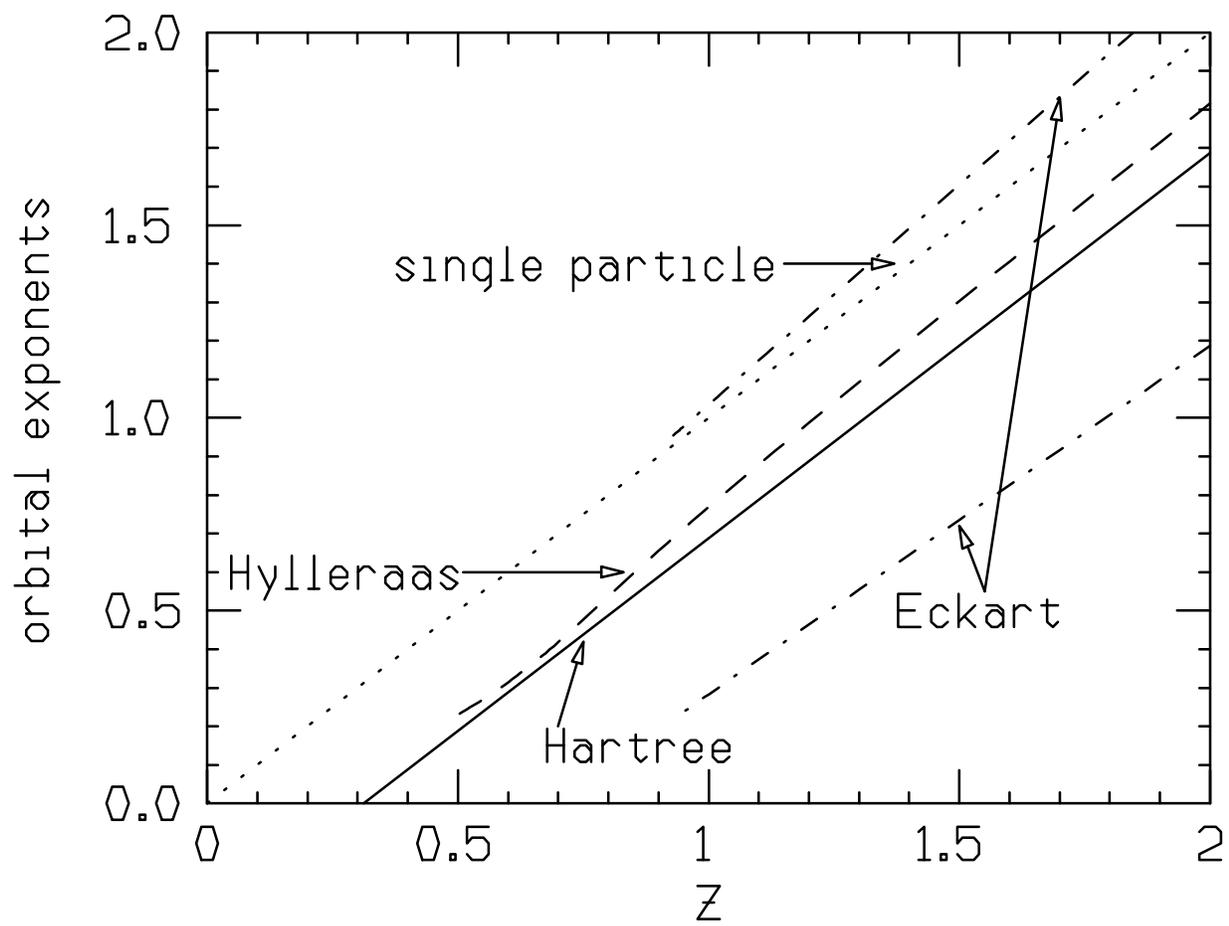
\begin{figure}
\caption { Orbital exponents for the two-electron atom versus ionic
charge $Z$ for the Hartree  (full line), Eckart (dot-dashed lines) and
Hylleraas (dashed line) wavefunctions. The dotted line shows the
orbital exponent for the one-electron atom ($=Z$).
}
\label{Fig. 1}
\end{figure}
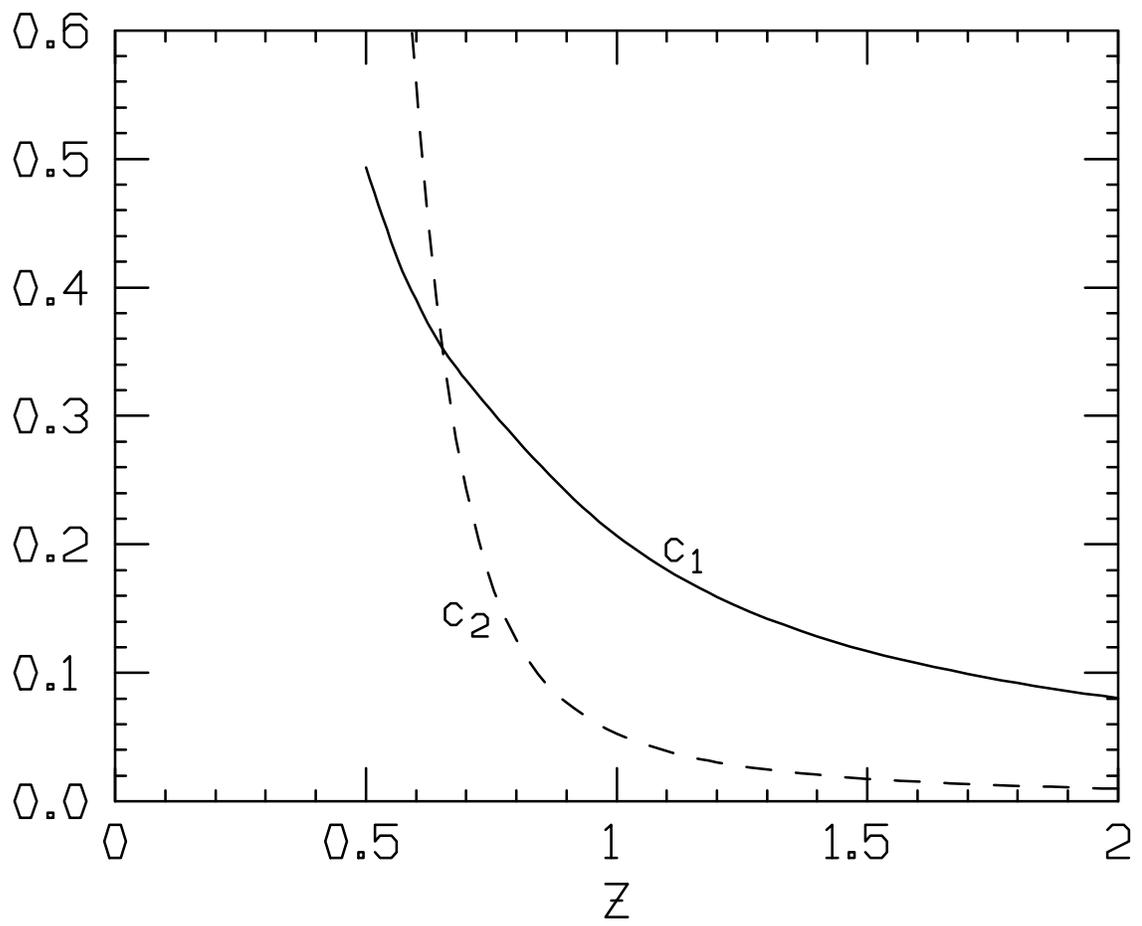
\begin{figure}
\caption { Parameters in the Hylleraas wavefunction Eq. (6) describing 
angular correlations ($c_1$, full line) and radial correlations ($c_2$, dashed line)
versus ionic charge
}
\label{Fig. 2}
\end{figure}
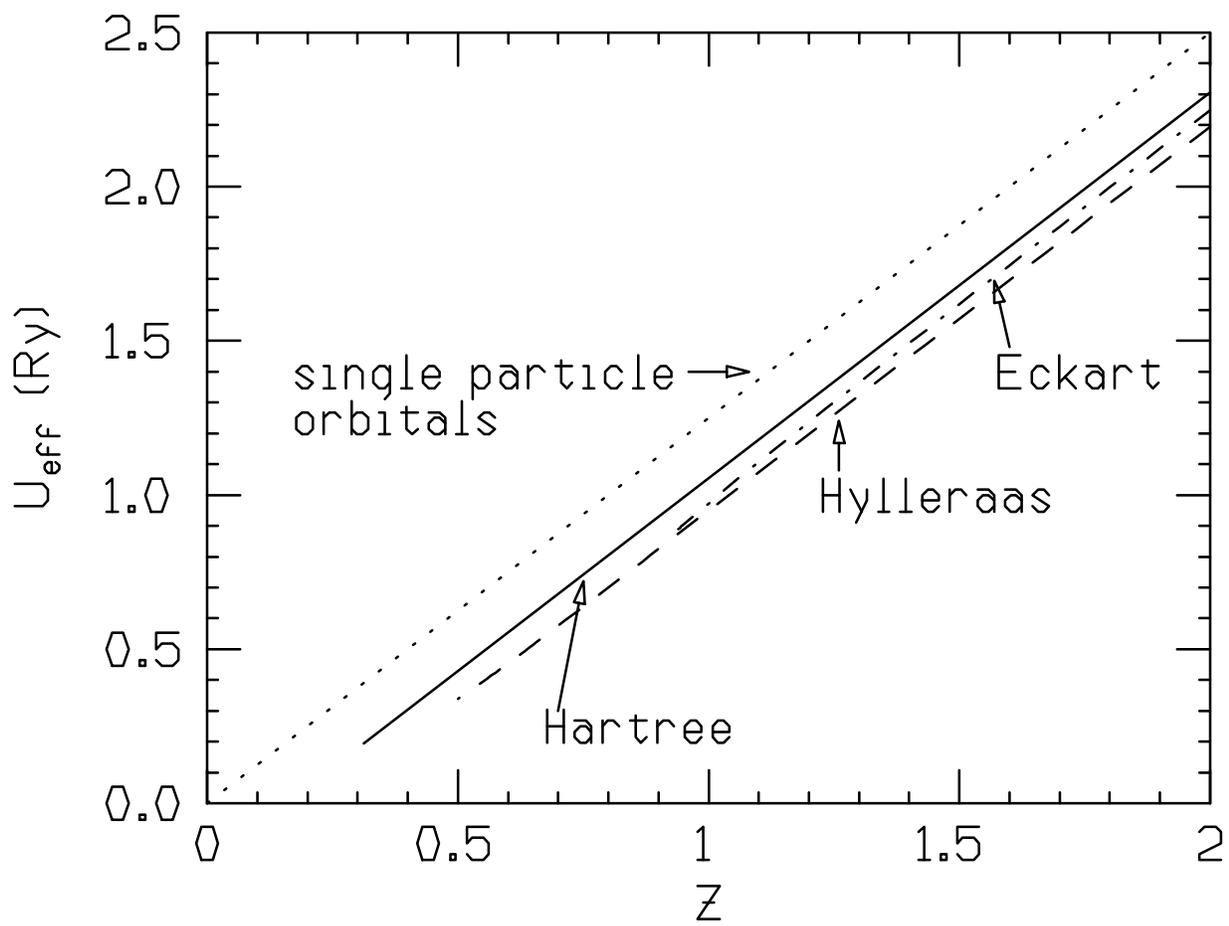
\begin{figure}
\caption { Effective $U$ (Eq. (7)) versus ionic charge $Z$ for the Hartree
(full line), Eckart (dot-dashed line) and Hylleraas (dashed line) wavefunctions.
The dotted line gives the bare $U$ (Eq. (8)) for the two electrons
in the single-particle orbital.
}
\label{Fig. 3}
\end{figure}
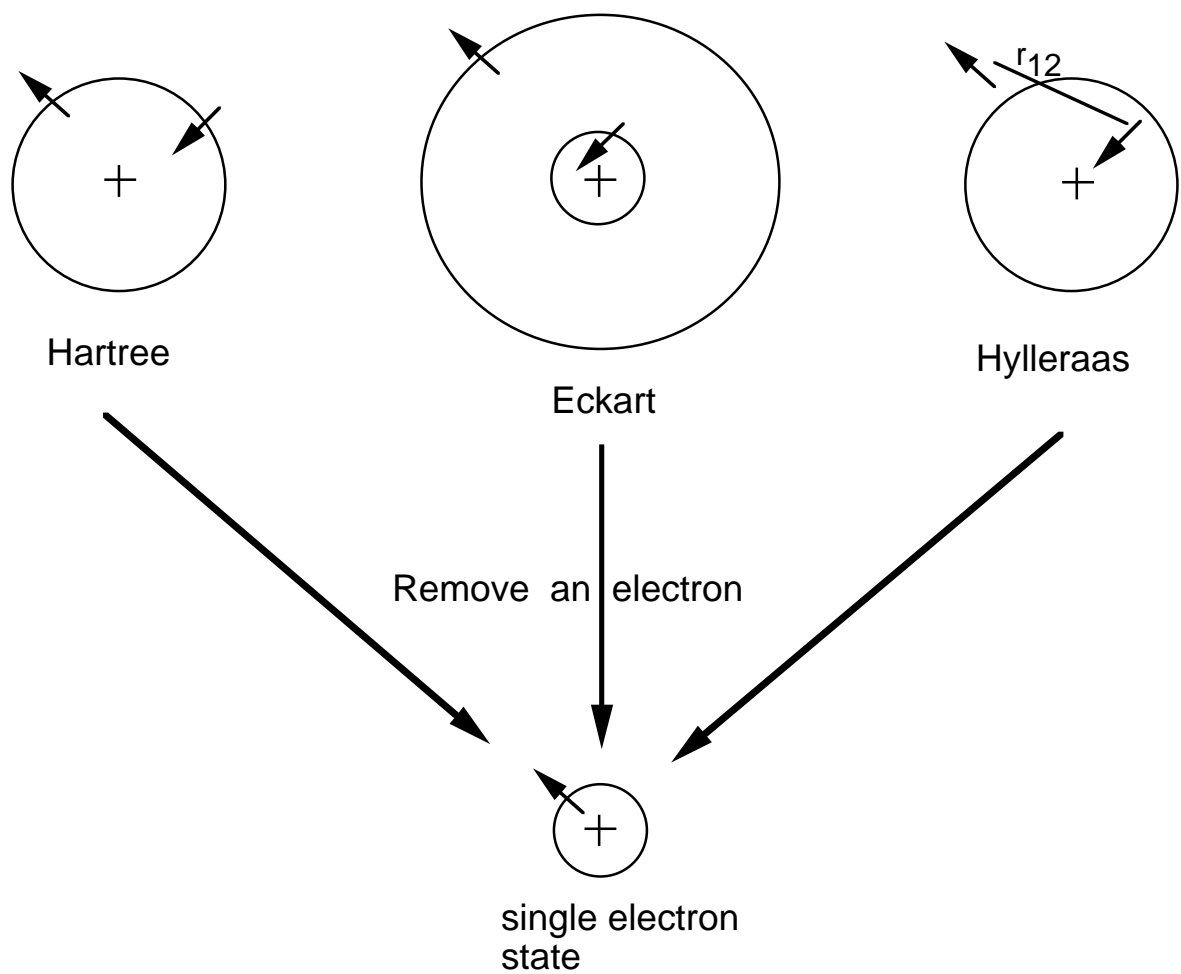
\begin{figure}
\caption { Qualitative depiction of the two-electron state: in the
Hartree wavefunction, both electrons occupy the same expanded orbital;
in the Eckart wavefunction, the electrons occupy different orbitals;
in the Hylleraas wavefunction, the amplitude of the wavefunction depends
on the relative angular and radial coordinates. When removing an electron
from any of these two-electron wavefunctions, the state of the remaining
electron has to change to become the eigenstate of the single-electron
atom.
}
\label{Fig. 4}
\end{figure}
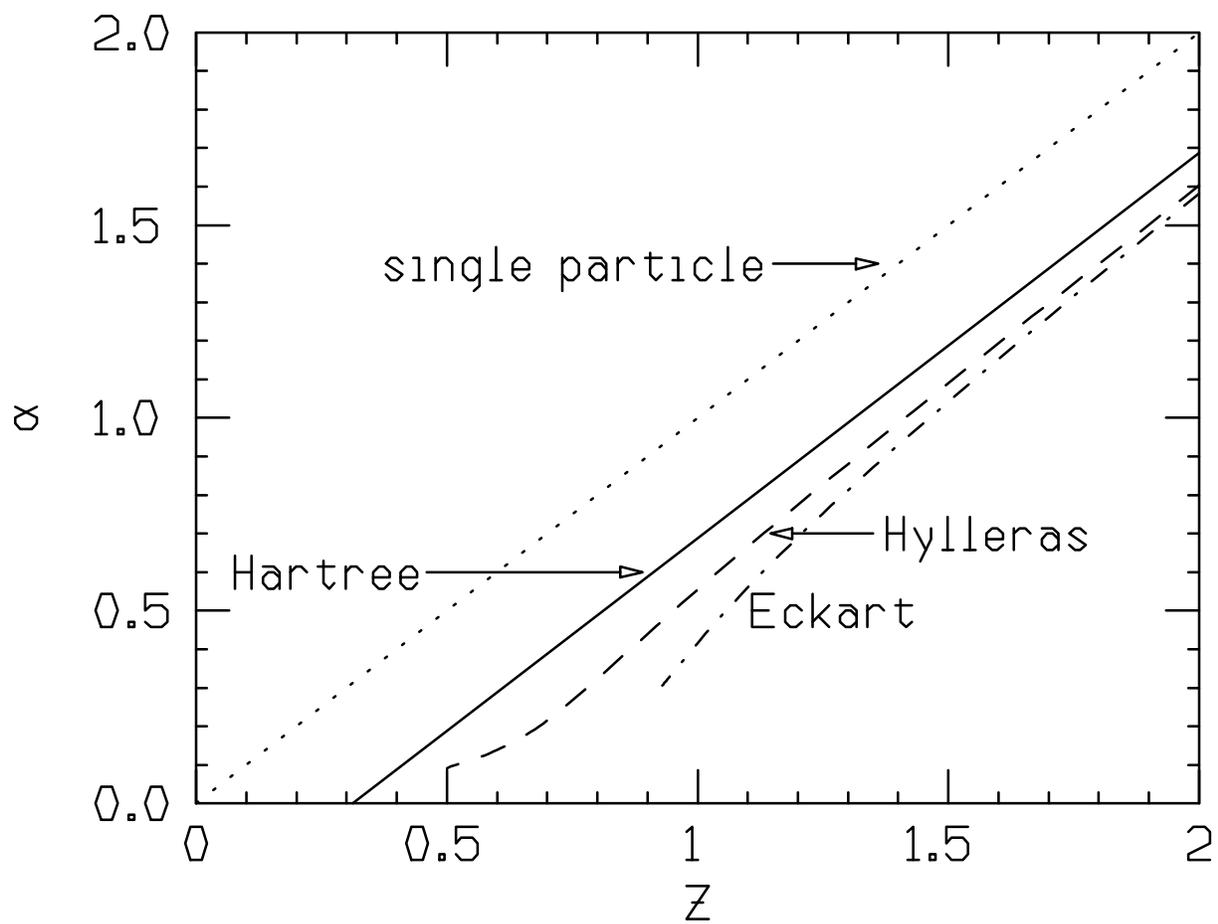
\begin{figure}
\caption { Orbital exponent of the 'hole wavefunction' $\varphi_\alpha$, defined
as the single particle wavefunction that yields maximal quasiparticle weight
for the hole spectral function Eq. (14), for the various approximations
to the two-electron wavefunction. The orbital exponent of the single electron
wavefunction is also shown (dotted line).
}
\label{Fig. 5}
\end{figure}
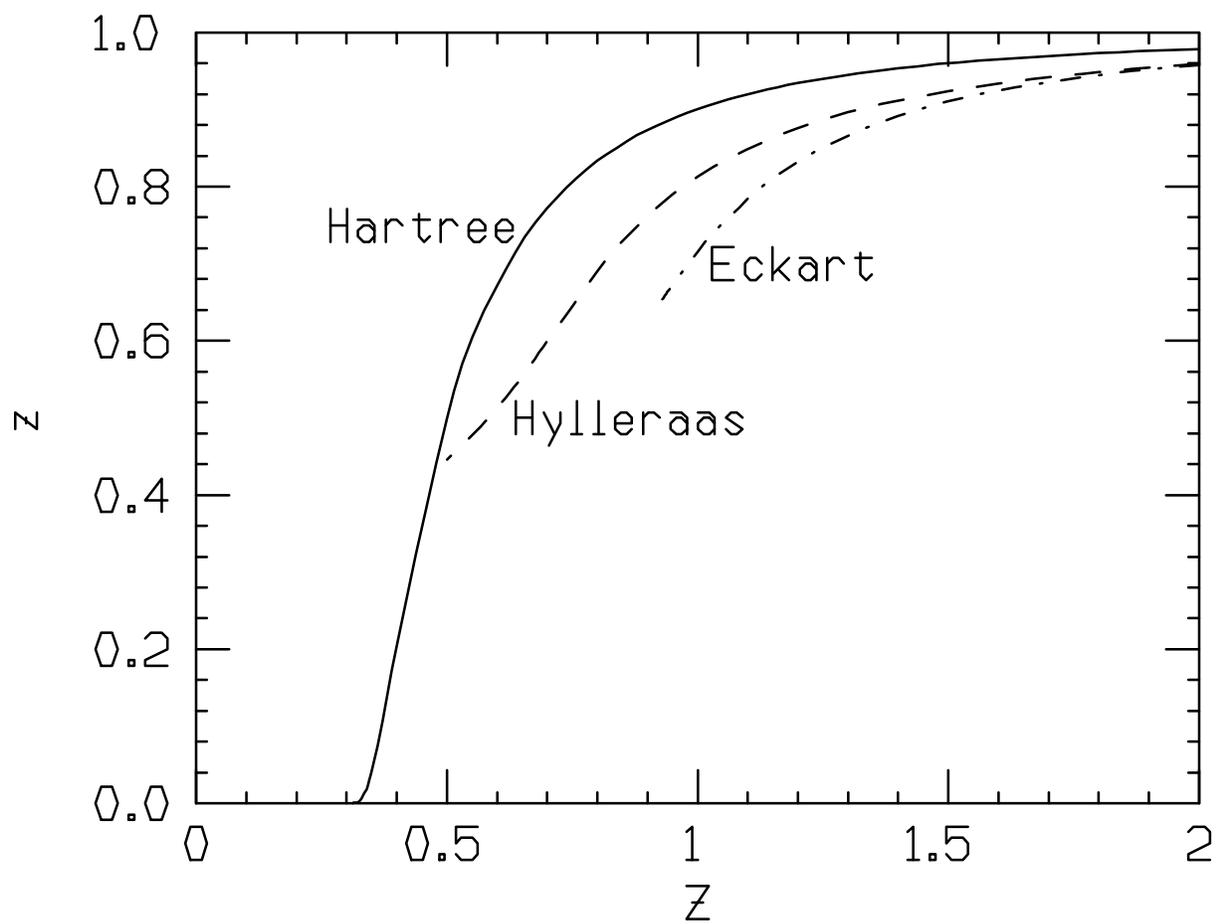
\begin{figure}
\caption { Quasiparticle weight for the hole, $z$, versus ionic charge $Z$,
for the various approximate two-electron wavefunctions. The difference
between these curves and $1$, the quasiparticle weight for the electron,
measures the importance of electron-hole asymmetry for given ionic charge.
}
\label{Fig. 6}
\end{figure}
\begin{figure}
\caption { Spectral function for hole creation for various values of the
ionic charge $Z$. The full lines indicate the magnitude of the coefficients of
the  $\delta$-functions within
the Hylleraas wavefunction, the dotted lines and symbols give the results for the
Hartree wavefunction. The 'width' of the $\delta-$functios is arbitrary.
As the ionic charge decreases the quasiparticle peak
($\delta$-function at $\omega=0$) decreases and the 'incoherent parts'
($\delta$-functions at negative frequencies) increase. At the same time
the energy of the incoherent excitations decreases in absolute value.
}
\label{Fig. 7}
\end{figure}
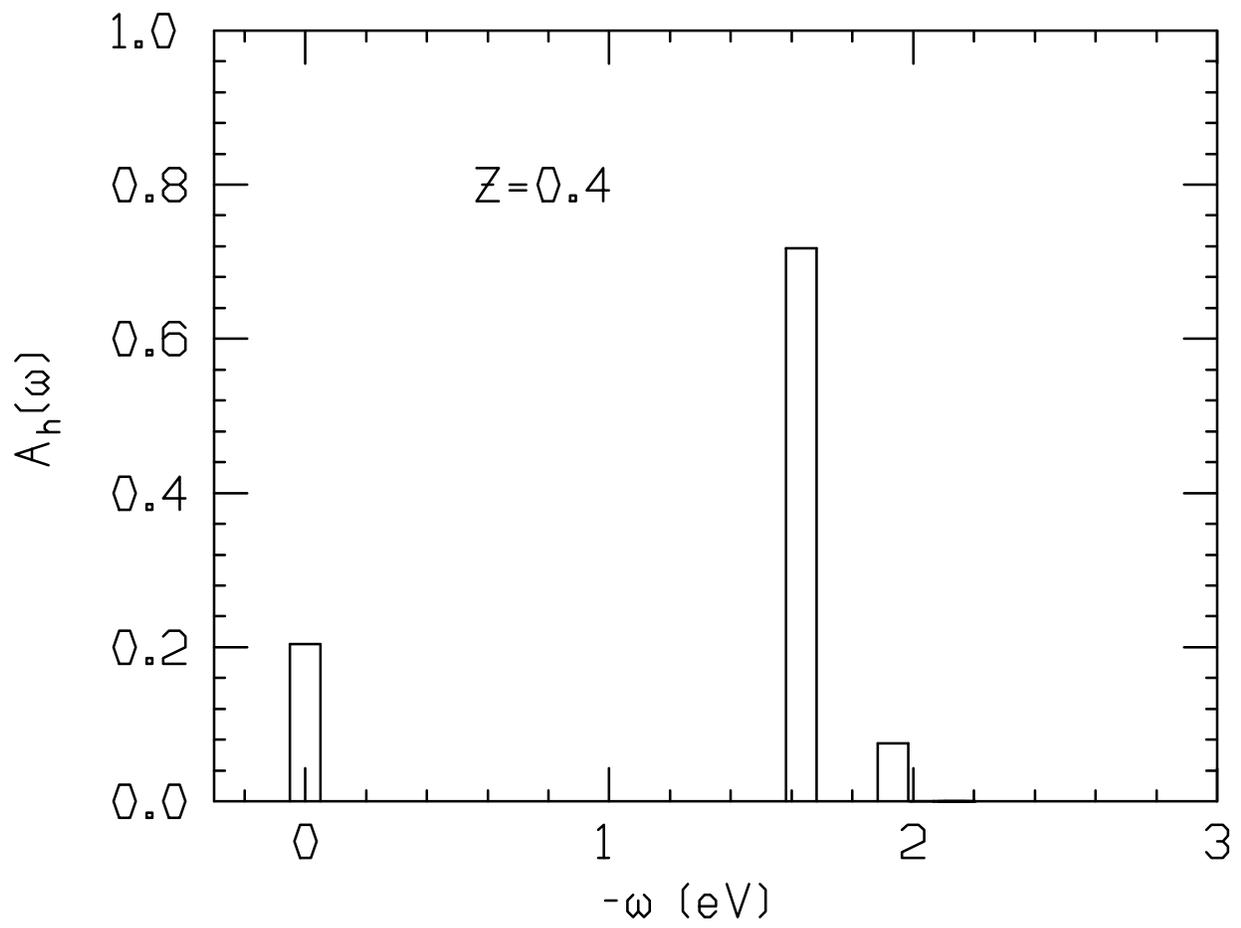
\begin{figure}
\caption { Spectral function for hole creation within the Hartree approximation
for ionic charge $Z=0.4$. Here the quasiparticle has smaller weight than
the incoherent excitations.
}
\label{Fig. 8}
\end{figure}
\begin{figure}
\caption { Spectral function for hole destruction (electron creation) in the
singly occupied atom within the Hylleraas approximation. Note that the energies
where the incoherent contributions appear are much lower for hole destruction
than for hole creation for the same $Z$. The quasiparticle peak has the same
height for hole destruction and hole creation. 
}
\label{Fig. 9}
\end{figure}
\begin{figure}
\caption { Dependence of the phenomenological one-electron spectral function on
band filling, for $\vec{k}$ the Fermi wavevector. 
Parameters used in Eqs. (43) and (46-51) are:
$S_1=0.5 $, $\epsilon_1 0.8=$, $\Gamma_1=0.2 $, $\epsilon_2 =0.5$, $\Gamma_2=0.2 $, $\alpha=0.5 $.
The zero frequency $\delta-$function is drawn with width $\Delta \omega=0.1$,
so that $A(k_F,\omega =0)=10eV^{-1}$ is the maximum height (for $z=1$).
As the band is filled with electrons (increasing $n$), the quasiparticle
weight $z(n)$ decreases and incoherent contributions appear, first at positive
energies (electron creation in the singly occupied atom) and at higher
$n$ at negative energies (electron destruction in the doubly occupied atom).
Electron creation in the empty atom and electron destruction in the singly occupied
atom give no incoherent contributions.
}
\label{Fig. 10}
\end{figure}
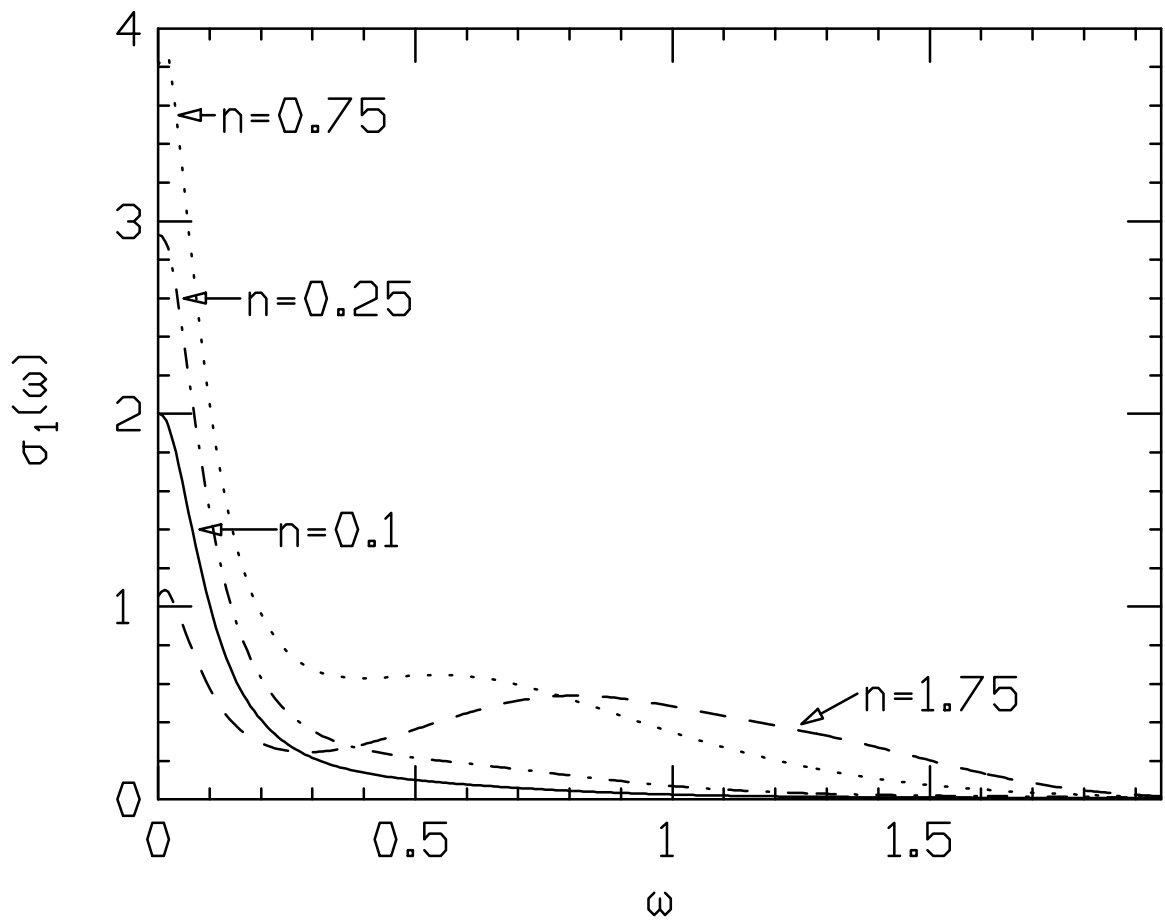
\begin{figure}
\caption { Frequency dependent conductivity $\sigma_1(\omega)$, Eq. (52), for
the parameters of the single electron spectral function used in Fig. 10.
As the band filling $n$ increases, spectral weight shifts from the low-frequency
Drude part to the high-frequency incoherent part. The zero-frequency conductivity
is high for electrons (small $n$) and low for holes (large $n$).
}
\label{Fig. 11}
\end{figure}

\widetext
\begin{table}
\caption{Orbital exponents and energy (in $Ry$) of two-electron atom 
for $Z=1$ ($H^-$) and
$Z=2$ ($He$) for the Hartree wavefunction Eq. (3) ($\bar Z$, $E_H$),
the Eckart wavefunction Eq. (4) ($Z_1$, $Z_2$, $E_E$) and
the Hylleraas wavefunction Eq. (6) ($Zk$, $E_{Hy}$). 
 The experimental value of the energy ($E_{exp})$ is also
given.}
\begin{tabular}{cllllllll}
$Z$ & $\bar Z $ & $E_H$ & $Z_1$ &
$Z _2$ & $E_E$ &  $Zk$ & $E_{Hy}$ &  $E_{exp}$ \cr
\tableline
1  & 0.6875  & -0.9453 & 1.0392 & 0.2832 & -1.0266  & 0.769 & -1.051 & -1.0554 \cr
2  & 1.6875  & -5.6953 & 2.1832 & 1.1885 & -5.7513  & 1.816 & -5.805  & -5.808 \cr
\end{tabular} \end{table}
 
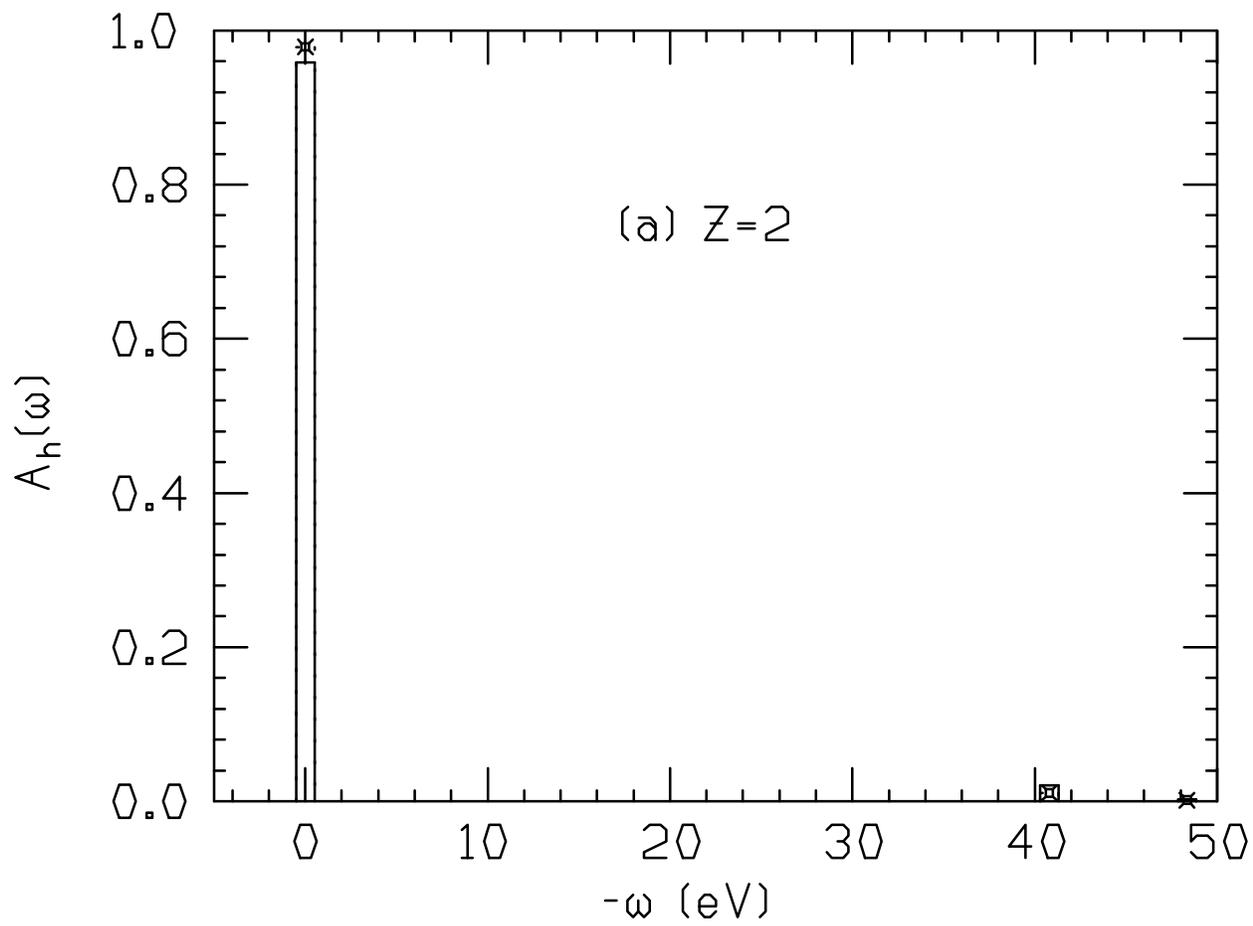
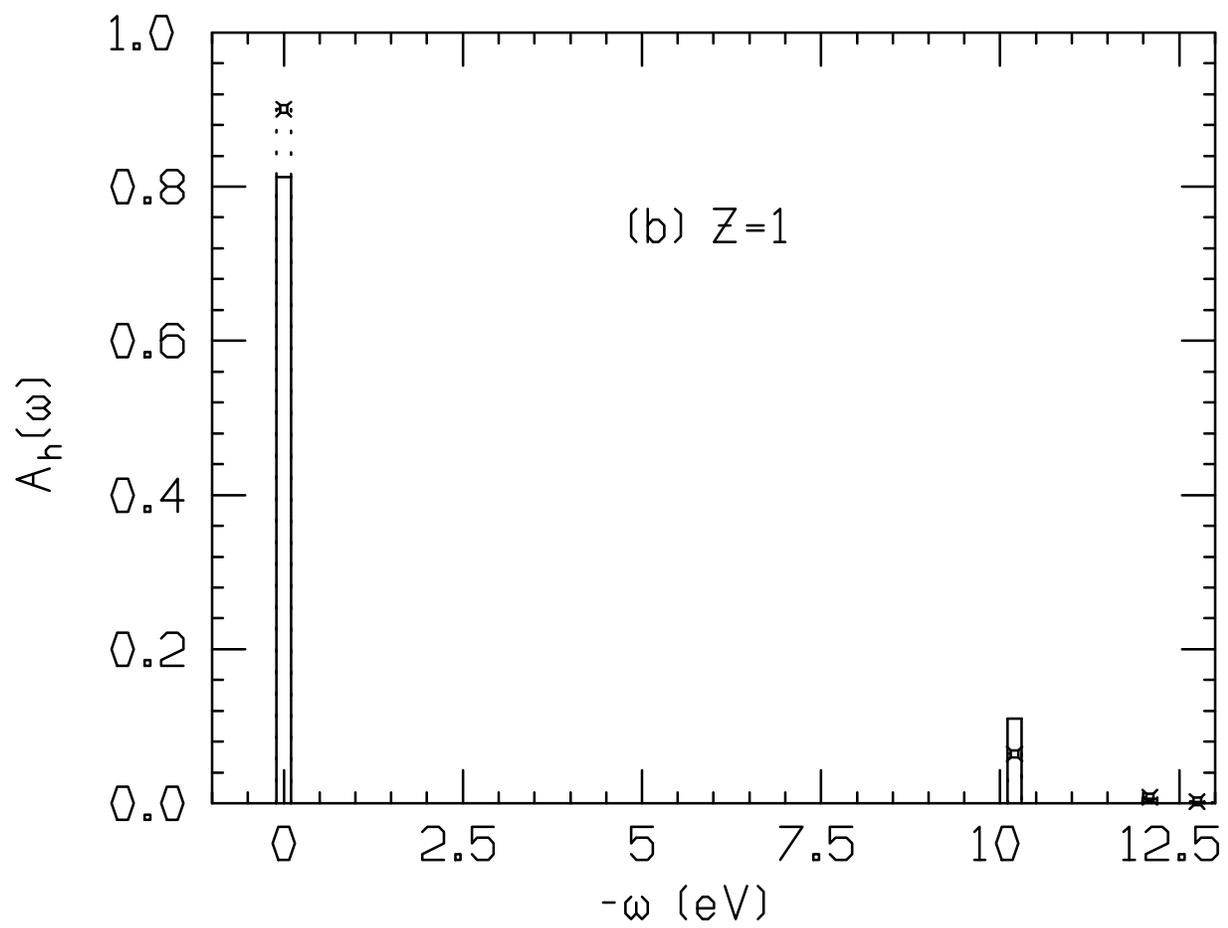
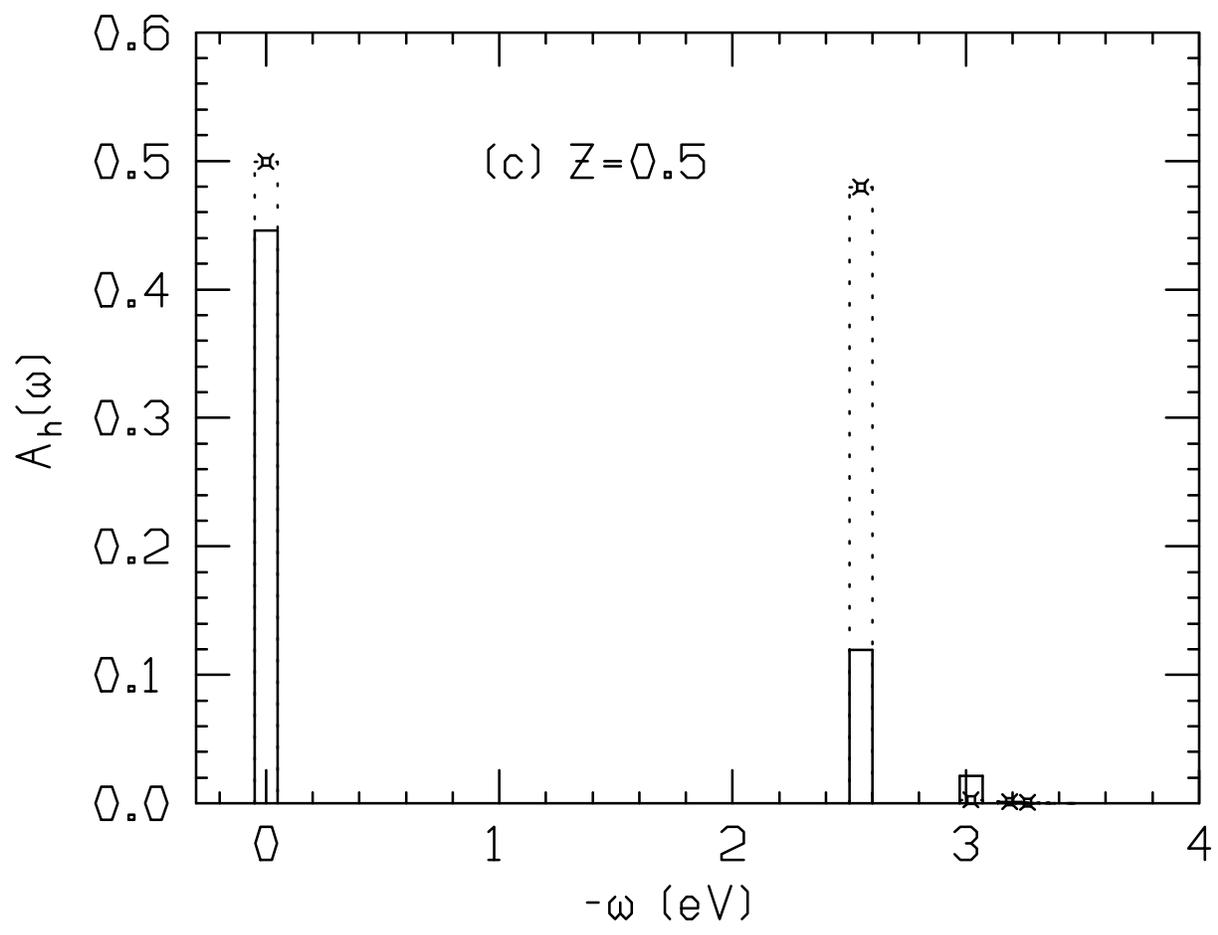
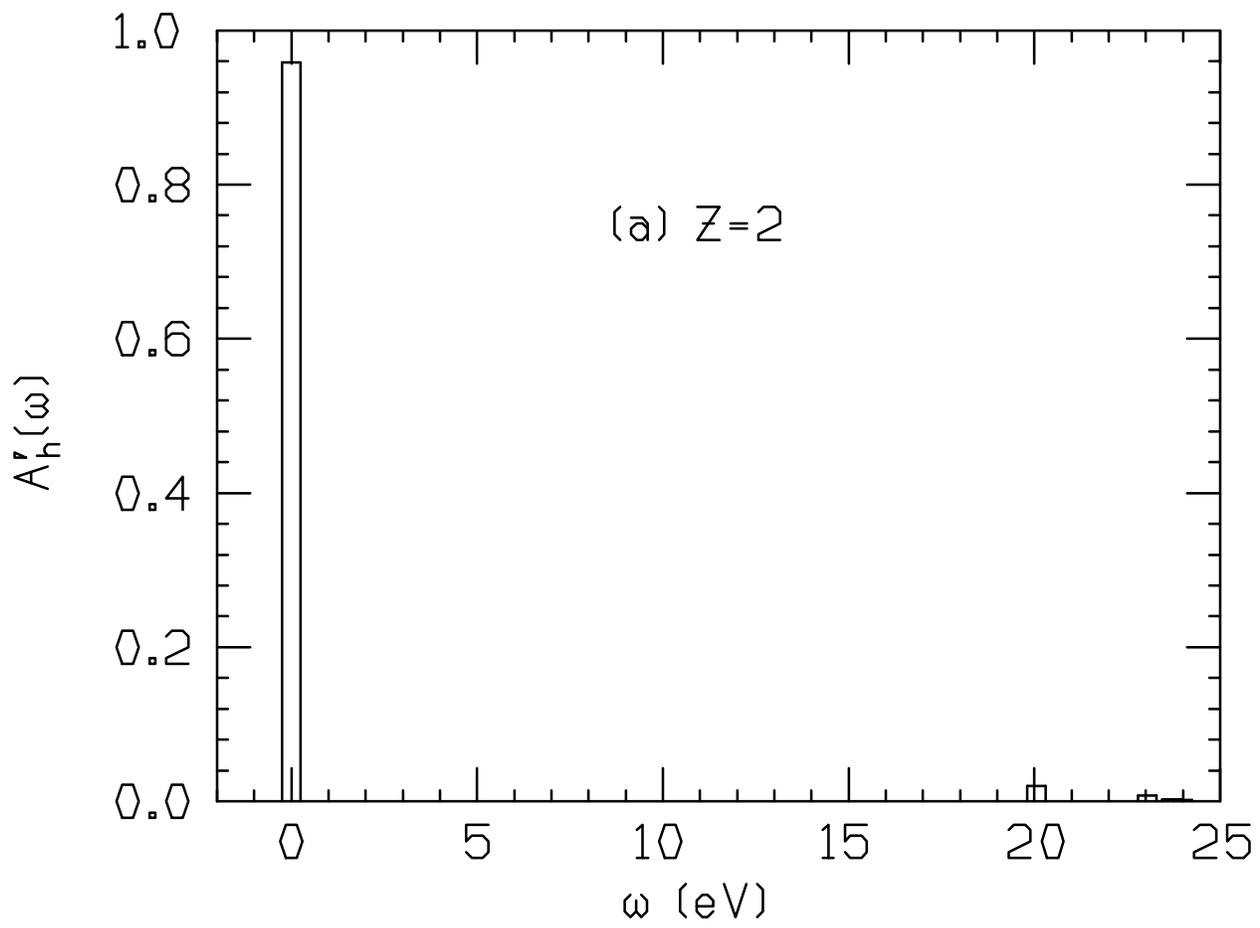
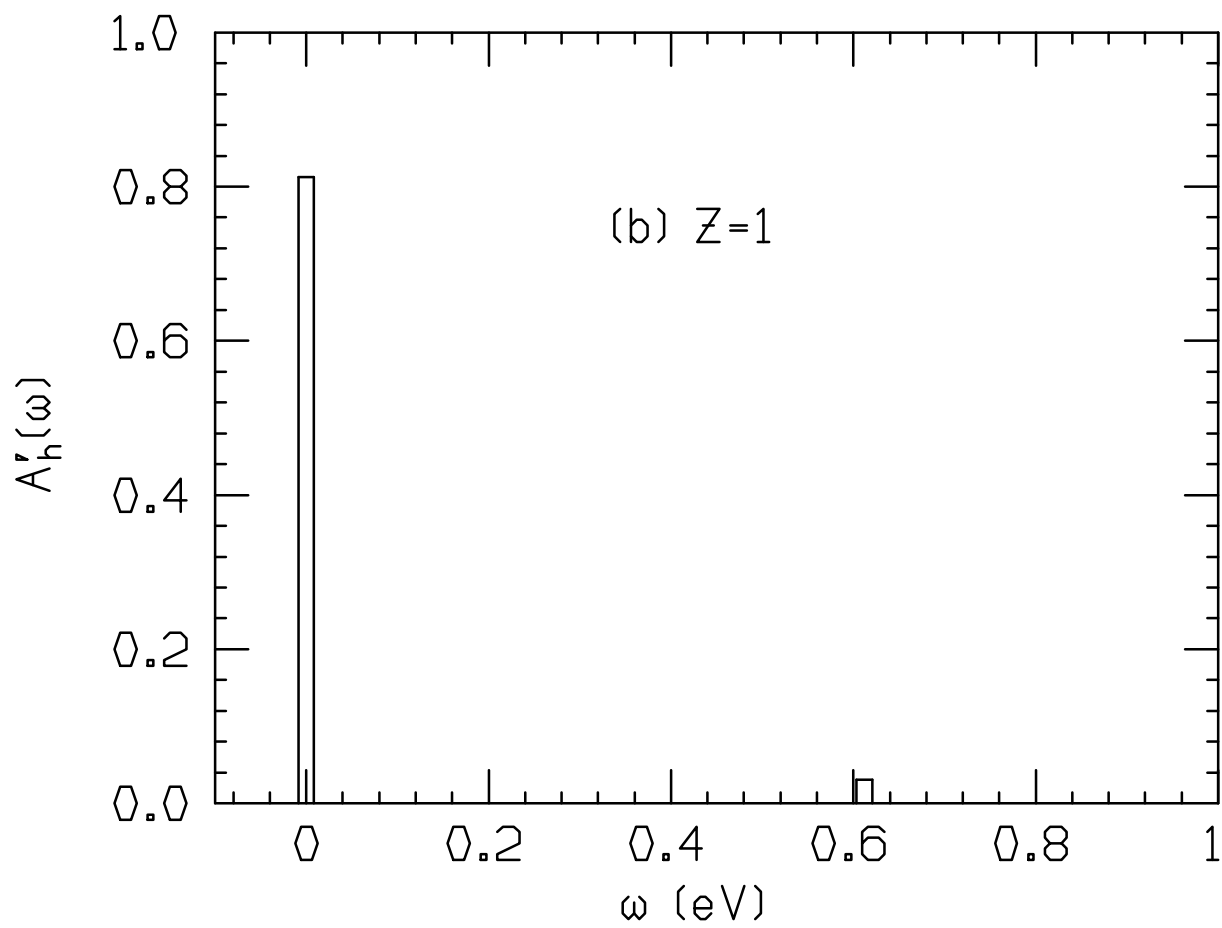
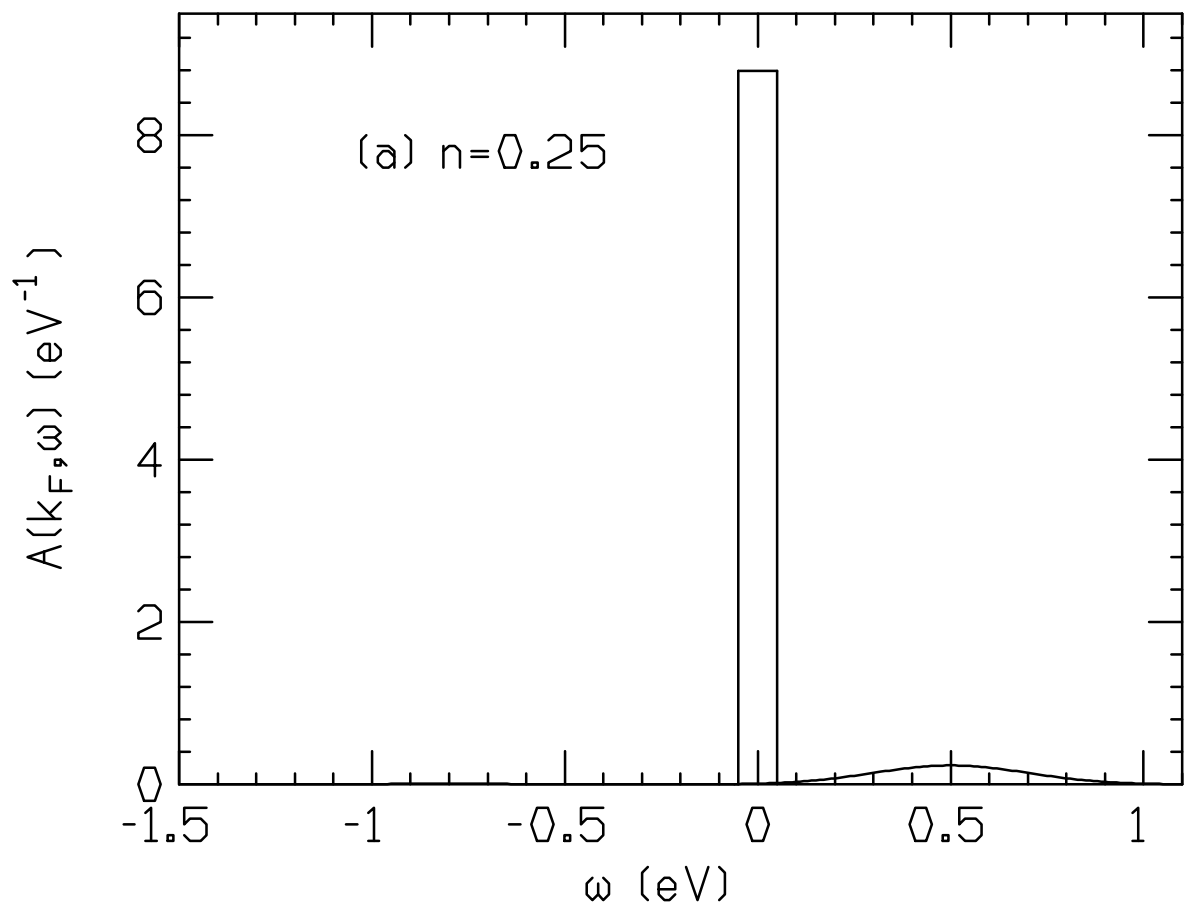
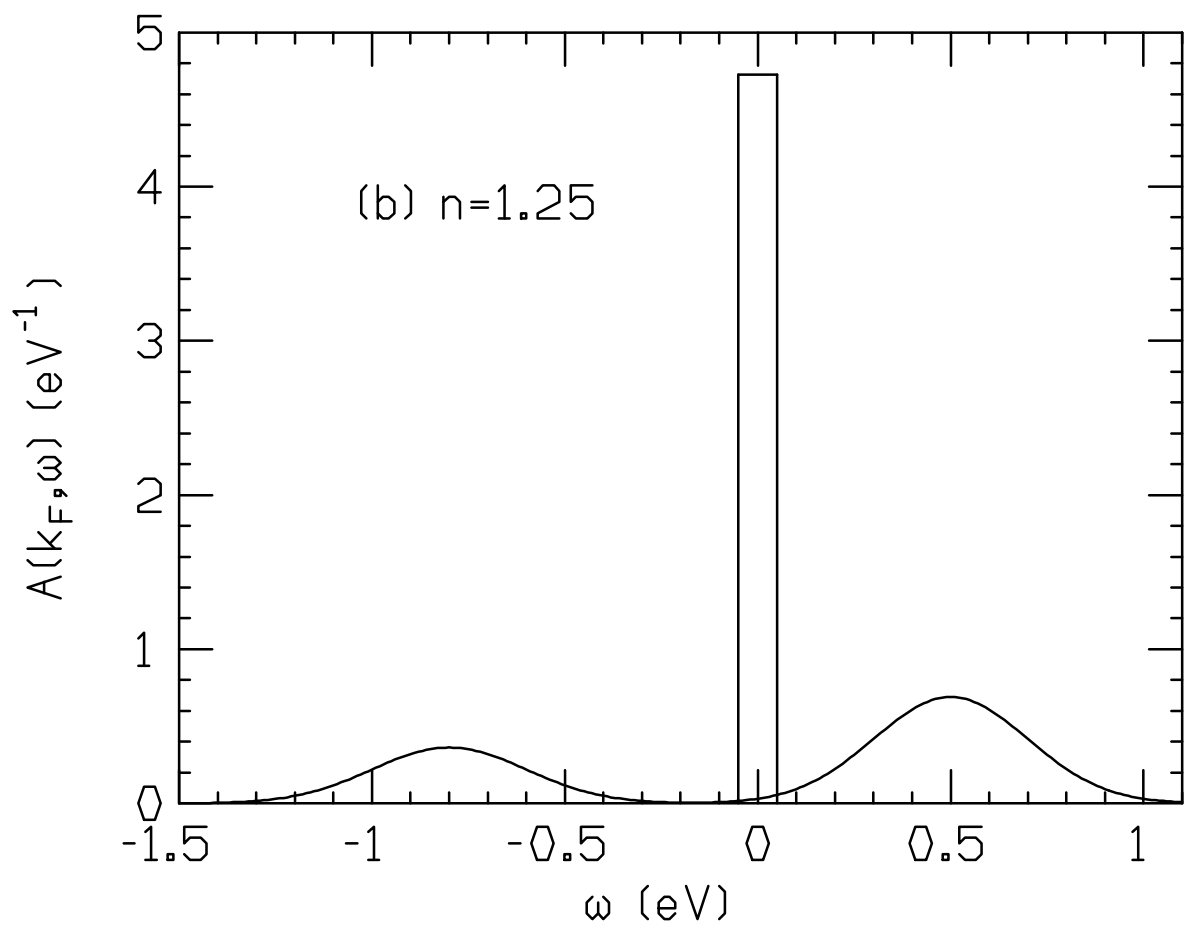
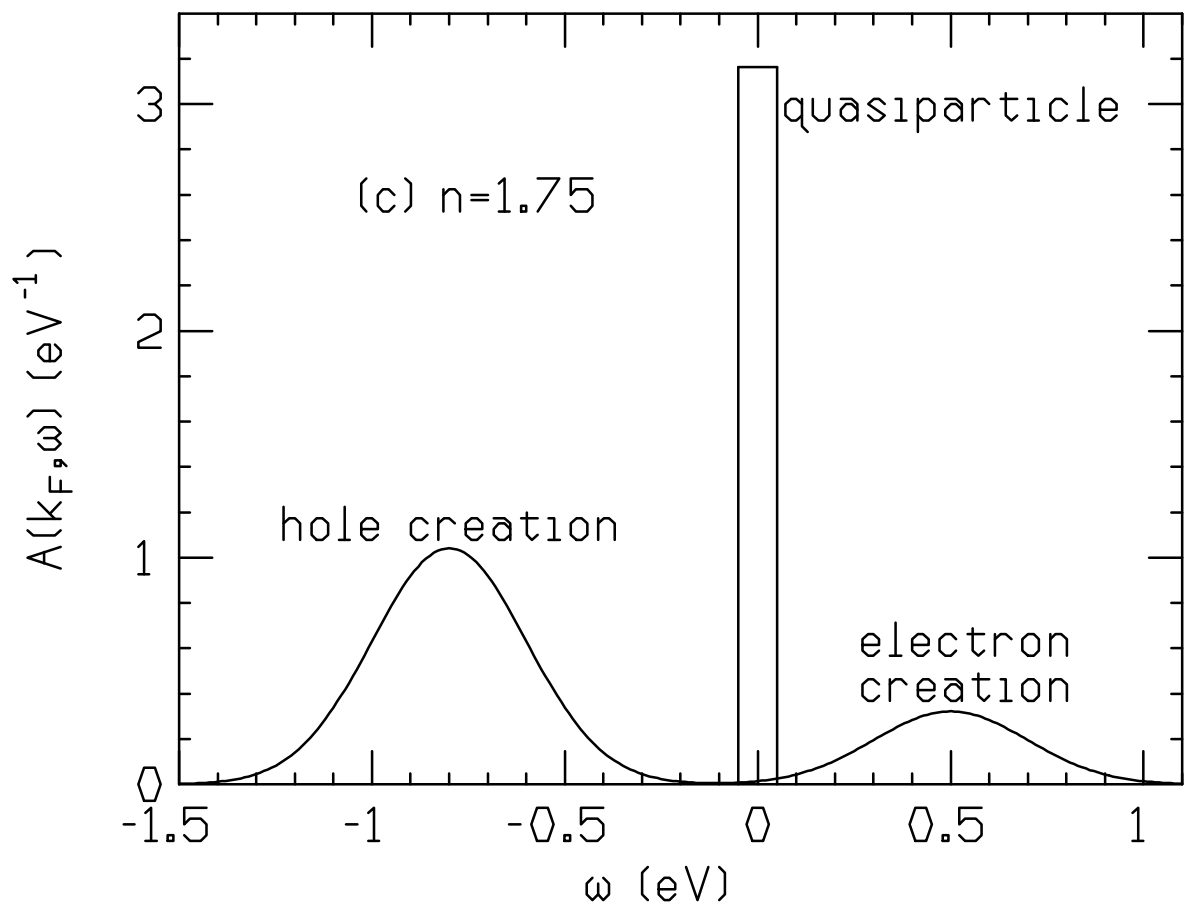
\end{document}